\documentclass[12pt]{iopart}

\usepackage{iopams}

\usepackage{amssymb}
\usepackage{amsthm}

\usepackage{latexsym} 

\usepackage{graphicx}
\usepackage{dcolumn}
\usepackage{bm}

\newcommand{\ket}[1]{|#1\rangle}
\newcommand{\braket}[2]{\langle #1|#2\rangle}
\newcommand{\hide}[1]{}
\newtheorem{theorem}{Theorem}
\newtheorem{lemma}{Lemma}

\newtheorem{proposition}{Proposition}

\newtheorem{requirement}{Requirement}
\newtheorem{claim}{Claim}
\newcommand{\ignore}[1]{}

\def\sM{{\cal M}}

\def\cM{{\cal M}}

\usepackage{latexsym}
\usepackage{graphicx}

\newcommand{\secref}[1]{Section~\ref{sec:#1}}
\newcommand{\thmref}[1]{Theorem~$\ref{thm:#1}$}

\newcommand{\lemref}[1]{Lemma~$\ref{lem:#1}$}

\renewcommand{\Pr}[1]{\mathbb{P}\left(#1\right)}
\newcommand{\Ex}[1]{\mathbb{E}\left[#1\right]}
\newcommand{\binom}[2]{\left(\begin{array}{c}#1 \\ #2\end{array}\right)}

\def\eps{\varepsilon}

\def\be{\begin{equation}}
\def\ee{\end{equation}}
\def\bea{\begin{eqnarray}}
\def\eea{\end{eqnarray}}

\newcommand{\ketbra}[1]{|\, #1\rangle\langle #1\,|}

\newcommand{\eqnref}[1]{(\ref{eq:#1})}

\def\X{{\cal{X}}}

\begin{document}

\title{Emergence of typical entanglement in two-party random processes}

\author{O.C.O. Dahlsten}
\ead{oscar.dahlsten@imperial.ac.uk}
\address{The Institute for Mathematical Sciences, Imperial College London, 53 Prince's Gate, South Kensington London, SW7 2PG, UK, and\\QOLS, Blackett
Laboratory, Imperial College London, London SW7 2BW, UK} 

\author{R. Oliveira}
\ead{rimfo@impa.br}
\address{Instituto Nacional de Matem\'{a}tica Pura e Aplicada - IMPA
Estrada Dona Castorina, 110 Jardim Bot\^{a}nico 22460-320, Rio de Janeiro, RJ - Brazil}
\author{M.B. Plenio}
\ead{m.plenio@imperial.ac.uk}
\address{The Institute for Mathematical Sciences, Imperial College London, 53 Prince's Gate, South Kensington London, SW7 2PG, UK, and\\QOLS, Blackett
Laboratory, Imperial College London, London SW7 2BW, UK}

\begin{abstract}
We investigate the entanglement within a system 
undergoing a random, local process. We find that 
there is initially a phase of very fast 
generation and spread of entanglement. 
At the end of this phase the entanglement
is typically maximal. In \cite{oliveira} we proved 
that the maximal entanglement is reached to a fixed 
arbitrary accuracy within $O(N^3)$ steps, where
$N$ is the total number of qubits. Here 
we provide a detailed and more pedagogical 
proof. We demonstrate that one can use the so-called 
stabilizer gates to simulate this process efficiently 
on a classical computer. Furthermore, we discuss 
three ways of identifying the transition from the phase 
of rapid spread of entanglement to the stationary 
phase: (i) the time when saturation of the maximal 
entanglement is achieved, (ii) the cut-off moment, when the
entanglement probability distribution is practically stationary,
 and (iii) the moment block entanglement 
scales exhibits volume scaling. We furthermore investigate 
the mixed state and multipartite setting. Numerically 
we find that classical and quantum correlations appear 
to behave similarly and that there is a well-behaved phase-space 
flow of entanglement properties towards an equilibrium, 
We describe how the emergence of typical entanglement can be 
used to create a much simpler tripartite entanglement 
description. The results form a bridge between certain 
abstract results concerning typical (also known as 
generic) entanglement relative to an unbiased distribution 
on pure states and the more physical picture of 
distributions emerging from random local interactions.

\end{abstract}


\maketitle


\section{Introduction}
Entanglement is a key resource in quantum information tasks
and therefore the exploration of the structure of entanglement
is of important concern in quantum information science \cite{Plenio V 05}.
Our quantitative understanding of this resource is very strong 
for bipartite entanglement; for reviews see 
\cite{Plenio V 05,Horodecki1,Wootters,Horodecki2,Plenio V 98,Eisert P 03}
, or refer to \cite{NC} for an introduction to quantum information 
tasks. However multipartite entanglement \cite{Plenio V 05,Eisert G 05} 
is much less well understood. In particular there appears to be 
a plethora of inequivalent classes of multipartite entanglement 
that are locally inequivalent 
\cite{Bennett MREGS,Linden PSW 99,Galvao PV 00,Wu Z 00,Ishizaka 04,Ishizaka P 05}. 
There is hope that one can cut down on this plethora by 
considering which classes are typical(generic) relative to a
certain measure on the set of states known as the unitarily 
invariant (Haar) measure \cite{hayden}. In that measure, 
practically all pure states of large numbers of spins are 
maximally entangled \cite{hayden,lubkin,lloyd,page,foong}. 
 This simplification suggests that the investigation
of generic entanglement may hold some promise. However,
an important question mark has existed as to whether 
this measure is physical, in the sense that it can be 
 approximated to arbitrary precision by two-particle 
interactions in a time that grows polynomially in the size 
of the system. The question has been raised in one form 
or another in \cite{oliveira,emerson1,emerson2,emerson3,smith}.

 Our key objective is to determine whether 
this is the case. We find and prove that it is indeed possible
to obtain generic entanglement properties in a polynomial 
number of steps in a two-party random process and give an 
explicit way of doing it. We also aim to gain a deeper 
understanding of the nature of the approach to the 
regime where entanglement displays generic behaviour. 
The paper both expands on the results of 
\cite{oliveira} and provides several new results.

The outline of the paper is as follows: We firstly discuss 
the key process that is used throughout this work: random
two qubit interactions. These are modelled as two-qubit 
gates on a quantum computer picked at random. We then 
give the results of the present work. Firstly we prove 
that the generic entanglement as well as purity is achieved 
efficiently. We additionally prove that one can use the 
so-called stabilizer gates to simulate this process 
efficiently on a classical computer in the sense
that the same averages will be achieved for relevant
quantities. We then discuss more 
in depth the observation that there is initially a 
phase of rapid spread of entanglement followed by a 
phase where the system is suffused with entanglement 
and the average entanglement across any bipartite cut 
is practically maximal. We discuss three ways of 
identifying the transition between these two phase: (i) the moment of saturation of the average 
entanglement, (ii) the cut off moment, and (iii) the 
moment the entanglement scales as the volume of the 
smaller of the two parties. We furthermore investigate 
the mixed state and multipartite setting.
We find numerically that the classical correlations 
appear to behave similarly and that there is a well-behaved 
phase-space flow to the attracting equilibrium entanglement, 
We describe how the emergence of typical entanglement 
can be used to create a much simpler tripartite entanglement 
description. Finally we give a conclusion as well as an 
extensive discussion of the future of this line of enquiry. 

The results form a bridge between certain abstract results 
 concerning typical (also known as generic) 
entanglement relative to an unbiased distribution on pure 
states with the more physical picture of entanglement
properties relative to distributions obtained by random 
local interactions. 

\section{Random two-party interactions}
Interactions in nature tend to be local two-party interactions. 
In the setting of qubits, that corresponds to two-qubit unitary 
gates. To have a concrete physical process in mind, consider a 
quantum computer which can perform arbitrary single 
qubit gates and CNOT \cite{gates} gates between any two qubits 
in the register. Let the randomisation process be the application 
of random gates on randomly chosen qubits. This process
leads to a distribution of pure states that will evolve 
over time gradually, and after a long time, approaching the 
flat distribution.

\subsection{The random walk}
\begin{figure}[th]
\centerline{
\includegraphics[width=7.5cm]{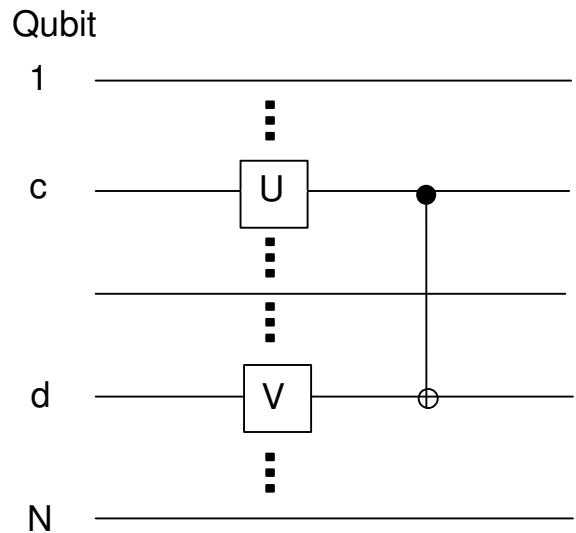}
}
\vspace{.0cm} \caption{Two-qubit random interactions. This shows steps (iii)
and (iv) of the random walk. For an explanation of such circuit diagrams see \cite{NC}. }
\label{fig:randomwalk}
\end{figure}

We shall discuss the evolution of states $\ket{\Psi}_Q$ in 
an N-qubit Hilbert space $Q=Q_1\otimes\ldots\otimes Q_N$ 
under a series of randomly chosen mappings. Each mapping is 
picked independently as in figure \ref{fig:randomwalk}.\\

\begin{enumerate}
\item Choose U and V independently from Haar measure on U(2)
(see appendix A for an introduction to the Haar measure). 
\item Choose a pair of distinct qubits $c$ and $d$, uniformly 
amongst all such pairs.
\item Apply $U\in U(2)$ to qubit $c$ and $V\in U(2)$ to $d$.
\item Apply a CNOT with target qubit $d$ and control 
qubit $c$.
\end{enumerate}

\subsection{Converges to uniform distribution}
The Markov Process described above converges to the uniform 
distribution on pure states (the distribution is described 
in Appendix A) because the CNOT and arbitrary single 
qubit unitaries are universal and the only probability measure
invariant under all unitaries is the Haar measure.
In the limit of infinitely many steps we have lost all bias 
towards being close to the initial state. In this setting 
any pure state is equally likely so we have the unitarily 
invariant distribution. 

It is important to note that the convergence rate to the 
final distribution is exponentially slow in the number of 
qubits since approximating an arbitrary unitary to a fixed 
precision using a set of fixed-size gates requires a number 
of steps that grows exponentially in the number of qubits
\cite{NC}. This leads one to question whether it is physically 
relevant to make statements relative to the unitarily 
invariant distribution. Interactions in nature tend to 
be two-body interactions and should therefore 
not get close to the unitarily invariant distribution 
in a feasible time, i.e. a time that scales polynomially
in the total number of qubits.

\subsection{Asymptotic Entanglement Distributions}
Consider the entanglement within a system undergoing the 
type of randomisation described above. We first give some 
facts about the asymptotic entanglement probability
distributions, and then give the key results of this 
work, which concern the rate and nature of approach to 
these asymptotic distributions.

When states are picked from the unitarily invariant measure 
there is an associated probability distribution of entanglement
 of a block of spins. Some of the first studies of 
this were \cite{lubkin,lloyd} and the explicit solution 
for the average entropy of entanglement ('Page's conjecture') was 
conjectured in \cite{page} and proven in \cite{foong}. It 
is given by
 \begin{equation}
\label{eq:page}
 \Ex{S(\rho_{A})}=\frac{1}{\ln 2}\left({
 \sum_{k=2^{N_B}+1}^{2^{N_A+N_B}}\frac{1}{k}-\frac{2^{N_A}-1}{2^{N_B+1}}
 }\right)
\end{equation}
with the convention that $N_A \leq N_B$ and where $N_A+N_B=N$, 
the total number of particles.

This can be used to show that the average entanglement is very
nearly maximal, i.e. close to $min(N_A,N_B)$, for large quantum 
systems where $N\gg 1$. It is also interesting to note that
there is a bound on the concentration of 
the distribution about this average. The probability that 
a randomly chosen state will have an entanglement $E$ that 
deviates by more than $\delta$ from the mean value $ \Ex{S(\rho_{A})}$ 
decreases exponentially with $\delta^2$\cite{hayden}. Therefore 
one is overwhelmingly likely to find a near-maximally entangled 
state if the system is large. 

\section{Theorem 1: Maximal average is achieved efficiently}
Despite the distribution on states requiring a number of 
steps that grows exponentially in the size of the system, 
we will now prove that achieving the entanglement 
distribution of generic quantum states to within any precision
 only requires a number of steps growing polynomially
in the size of the system. 
An intuition why this is possible can be gained by noting 
that many states have the same entanglement, so attaching 
an entanglement value to all states results in a 'coarse-grained' 
state space which can be sampled in fewer steps. 
 We now state our main
\begin{theorem}\label{thm:main}Suppose that $N_B-N_A=t\geq 0$ and some $\eps\in (0,1)$ is given. Then if the number of steps $n$ satisfies
$$n\geq \frac{9N(N-1)[(2\ln 2)N + \ln\eps^{-1}]}{4},$$ 
we have 
\begin{eqnarray}\label{eq:entropybound}\Ex{S(\rho_{A,n})} &\geq & N_A - \frac{2^{-t} + \eps}{\ln 2}\\
\label{eq:maingoal}\Ex{\max\limits_{\ket{\Psi}_{AB}\mbox{\small max. entangled}}\braket{\psi_n}{\Psi}}&\geq &1 - \sqrt{\frac{4(2^{-t}+\eps)}{\ln 2}}.\end{eqnarray}\end{theorem}
Notice that the second statement of the Theorem is most 
relevant for $t\gg 1$. This is because maximal entanglement $N_A$ is not exactly
achieved when $N_B-N_A=O(1)$. 
A similar problem is present in the asymptotic case 
analysis of \cite{hayden}.

\subsection{Guide to the proof of Theorem 1}
Firstly we simplify the problem by noting that the purity 
$Tr(\rho_{A,n}^2)$ bounds the entanglement very tightly 
 in the regime of almost maximal entanglement. 
Purity is more convenient to deal with so we study the 
convergence rate of the purity to its asymptotic value. 
We expand the global density matrix in terms of elements 
of the Pauli group and track the time evolution of the 
coefficients. It will turn out that the computation 
of $Tr\rho^2$ requires only the knowledge of the squares
of some of these coefficients. It is then a key innovative
step to realize that the relevant coefficients evolve 
according to a Markov Chain on a small state 
space which we give and prove explicitly. We then 
use relatively recent Markov Chain convergence rate 
analysis tools to determine how fast this chain converges 
to its stationary distribution. These arguments center 
around the size of the 'spectral gap' of the stochastic 
matrix $P$ defining the Markov Chain, which simply means 
the difference between the second largest eigenvalue 
$\lambda_1$ and the eigenvalue $1$. This is 
because $P^k=S {\rm diag}(1,\lambda_1^k,\lambda_2^k...)S^{-1}$ 
for some matrix $S$ so the $\lambda_1^k$ will define 
the most slowly decaying term and thus govern the
distance to the stationary distribution.

\subsection{Using purity to bound entanglement}
Most of our mathematical work will be in estimating the quantity 
$\Ex{Tr(\rho_{A,n}^2)}$. More precisely, \thmref{main} follows 
from
\begin{lemma}\label{lem:main}
For all $n$, 
$$\left|\Ex{Tr(\rho_{A,n}^2)} - \frac{2^{N_A}+2^{N_B}}{2^N+1}\right|\leq 4^{N}\exp\left(-\frac{4n}{9N(N-1)}\right).$$
\end{lemma}
 Before we prove Lemma 1 we demonstrate that Theorem 1 may
be deduced from it. To see the implication, first notice 
that the Von-Neumann entropy $S(\rho_{A,n})$ is lower-bounded 
by the {\em R\'{e}nyi entropy}
$$S_2(\rho_{A,n}) = -\log_2(Tr(\rho_{A,n}^2)).$$
By the concavity of $\log_2$, we have 
$\Ex{S_2(\rho_{A,n})}\geq -\log_2(\Ex{Tr(\rho_{A,n}^2)})$, 
and plugging in a $n$ as suggested in the Theorem,
$$\Ex{S_2(\rho_{A,n})}\geq -\log_2\left(2^{-N_A}\frac{1+2^{-t}}{1+2^{-N}} + 2^{-N_A}\eps\right)\geq N_A - \log_2(1+2^{-t}+\eps)\geq N_A - \frac{2^{-t}+\eps}{\ln 2}.$$ For the second statement, we note that by Uhlmann's Theorem, the expectation in the LHS of \eqnref{maingoal} is given by a fidelity 
of reduced density matrices. We can then use well-known relationships between distance measures on density matrices \cite{NC} to deduce
\begin{eqnarray}\label{eq:expfidelity}
    1-\Ex{\max\limits_{\ket{\Psi}_{AB}\mbox{\small max. entangled}}
    \braket{\psi_n}{\Psi}}&=&\Ex{1-F(I_A/2^{N_A},\rho_{A,n})}\\ 
    &\leq & \sqrt{\Ex{[1-F(I_A/2^{N_A},\rho_{A,n})]^2}}\\
    &\leq & \sqrt{\Ex{\|I_A/2^{N_A}-\rho_{A,n}\|_{tr}}}\\ 
    &\leq &\sqrt{2 \Ex{D(\rho_{A,n}||I_A/2^{N_A})}}.\end{eqnarray}
Here $D(\sigma||\rho):=Tr[\sigma\log_2\sigma-\sigma\log_2\rho]$ is the relative entropy distance, which in this particular case 
reads 
$$D(\rho_{A,n}||I_A/2^{N_A}) = N_A - S(\rho_{A,n}).$$
Using the same argument as above, 
\begin{eqnarray*}
    D(\rho_{A,n}||I_A/2^{N_A}) &\leq & 
    N_A - \Ex{S_2(\rho_{A,n})}\leq N_A +\log_2(\Ex{Tr[\rho_{A,n}^2]})\\
    &\leq & N_A + \log_2\left(\frac{2^{N_A}+2^{N_B}}{2^{N}+1} + 8^{N}\exp\left(-\frac{4n}{9N(N-1)}\right)\right),
\end{eqnarray*}
and Theorem 1 follows.

We are now proceeding with the proof of Lemma 1. The explanation
of the first main ingredient will require some basic tools from
linear algebra.
\subsubsection{Linear algebra in the space of Hermitian operators}

The proof of \lemref{main} takes an indirect route that requires 
a quick detour into linear algebra.

Let us make the following conventions: $V[c]$ represents an 
operator $V$ acting on qubit $c$. Denote the Pauli-operators (defined in appendix C) 
by $\sigma_i$ for $\{ i=1,2,3 \}$ and use $\sigma_0={\bf 1}$.
For a string $p=p_1\dots p_N\in \{0,x,y,z\}^N$,
$$\Sigma_p = 2^{-N/2}\bigotimes_{i=1}^N\sigma_{p_i}[i]$$
is a tensor product of Pauli operators, normalized so that 
$Tr[\Sigma^2_p]=1$.
It is well-known that the operators $\Sigma_p$ form an 
orthonormal basis of the real vector space of Hermitian 
matrices over $N$ qubits, with the inner product between 
$A$ and $B$ given by $Tr(AB)$. Therefore, if we write
$H = \sum_{p}h(p)\Sigma_p$ for a Hermitian operator $H$, 
we have 
$$Tr[H^2] = \sum_{p}h(p)^2.$$
Let us also note that if $B\subset [1,\ldots,N]$ is a 
non-empty subset of qubits 
and $A=[1,\ldots,N]\backslash B\neq \emptyset$, we can express the tracing out 
of $B$ in the following form
\begin{eqnarray*}
    Tr_B[H] &=& \sum_{p\in \{0,x,y,z\}^N}h(p)Tr_B[\Sigma_p]\\
    &=& \sum_{p\in \{0,x,y,z\}^N}h(p) \left(\frac{\prod_{j\in B}
    Tr[\sigma_{p_j}[j]]}{2^{|B|/2}}\right)\times 
    \left(\frac{\bigotimes_{i\in A}\sigma_{p_i}[i]}{2^{|A|/2}}\right)\\
    &=& 2^{|B|/2}\sum_{p\in \{0,x,y,z\}^N\,:\, \forall j\in B, p_j=0}h(p)
    \frac{\bigotimes_{i\in A}\sigma_{p_i}[i]}{2^{|A|/2}}.
\end{eqnarray*}
\subsubsection{Back to our problem}

Let us now discuss how to apply the above to our problem. Assume that we write
$$\ketbra{\psi_{n}} \equiv \sum_{p\in\{0,x,y,z\}^N} \xi_n(p)\Sigma(p),$$
where the $\xi_n(p)$'s are real coefficients. Then
\begin{eqnarray}
    \nonumber 
    Tr(\rho_{A,n}^2) &=& Tr\left[\left(2^{|B|/2}
    \sum_{p\in\{0,x,y,z\}^N\,:\,\forall j\not\in A p_j=0}
     \xi_n(p)\frac{\bigotimes_{i\in A}\sigma_{p_i}[i]}{2^{|A|/2}}
     \right)^2\right]\\ 
     \label{eq:fundamental} &=& 2^{N_B}\sum_{p\in\{0,x,y,z\}^N
     \,:\,\forall j\not\in A, p_j=0} \xi_n(p)^2. 
\end{eqnarray}
Thus it suffices to determine the evolution of the positive
coefficients $\xi_n(p)^2$. Our key idea is to map this evolution to a classical Markov chain and use a rapid mixing analysis to understand this evolution.

\subsection{Evolution of the coefficients}

 The main and final goal of this section is to map the evolution of 
the coefficients $\xi_n^2(p)$ under the random applications 
of quantum gates onto a random walk over states $p\in\{0,x,y,z\}^N$.
 
To achieve this end, several preliminary calculations are necessary. 
However, we will only use the following assumption about $U$ and $V$:

\begin{requirement}\label{assump:rand}At each step of the process, 
$U$ and $V$ are independently chosen from a measure on $U(2)$ such 
that, if $T$ is distributed according to the same measure, $F=F(A,B)$ 
is a bi-linear function of $A$ and $B$, and $a,b\in\{0,x,y,z\}$,
$$\Ex{F(T\sigma_aT^\dag,T \sigma_b T^\dag)} = 
\left\{\begin{array}{ll}F(I,I),& a=b=0;\\ 
\frac{1}{3}\sum_{w\in\{x,y,z\}}F(\sigma_w,\sigma_w), & 
a=b\neq 0\\ 0, &\mbox{otherwise.}\end{array}\right.$$
\end{requirement}

It is shown in \secref{assumption} that Haar measure on 
$U(2)$ does have the above properties, and in section 
\secref{stabilizers} that the set of single qubit 'stabilizer gates' 
also does.

\subsubsection{Basic aspects}

Now suppose we are given $\ket{\psi_n}_{AB}$ and choices of 
$c$, $t$, $U$ and $V$. Then 
$\ket{\psi_{n+1}}=I_{[1,\ldots,N]\backslash\{c,t\}}\otimes W_n\ket{\psi_n}$, where
$$W_n = CNOT[c,t] (U[c]\otimes V[t]).$$
Therefore,
$$\ketbra{\psi_{n+1}} = \sum_{q\in\{0,x,y,z\}^N}\xi_n(q) (I_{[1,\ldots,N]\backslash\{c,t\}}\otimes W_n)\Sigma_q (I_{[1,\ldots,N]\backslash\{c,t\}}\otimes W_n)^\dag $$
and for any $p\in\{0,x,y,z\}^N$ we find
\begin{eqnarray}
    \xi_{n+1}(p) &=& tr[\Sigma(p)|\psi_{n+1}\rangle\langle
    \psi_{n+1}|]\nonumber \\
    &=& \frac{1}{4}\sum_{q:\forall i\not\in\{c,t\} 
    q_i=p_i} \xi_n(q) Tr[(\sigma_{p_c}[c]\sigma_{p_t}[t])
    W_n(\sigma_{q_c}[c]\sigma_{q_t}[t]) W_n^\dag] 
\end{eqnarray}
and
\begin{eqnarray}
    \xi_{n+1}^2(p)&=&
    \frac{1}{16}\sum_{q,q':\forall i\not\in\{c,t\} q_i=q'_i=p_i}
    \xi_n(q)\xi_n(q') G_n(p,q,q')
\end{eqnarray}
where
$$G_n(p,q,q')\equiv Tr\left[(\sigma_{p_c}[c]\sigma_{p_t}[t])
W_n(\sigma_{q_c}[c]\sigma_{q_t}[t]) W_n^\dag\right] 
Tr\left[(\sigma_{p_c}[c]\sigma_{p_t}[t])W_n(\sigma_{q'_c}[c]
\sigma_{q'_t}[t]) W_n^\dag\right].$$
The above expression would appear to suggest that $\xi^2(p)$ 
depends on the non-positive $\xi_n,\xi'_n$ which would prevent
the formulation of a Markov process for $\xi^2$. However, we are
only interested in averages and this will be the key for a
further simplification.
Let us consider the expectation of $G_n(p,q,q')$
conditioned on the values of $c$, $t$ and $\psi_n$. Let 
us notice first that $G_n(p,q,q')$ can be rewritten as
\begin{eqnarray}\label{eq:fourtraces}
    \hspace*{-1.cm}G_n(p,q,q')=Tr[\sigma_{\hat{p}_c}U\sigma_{q_c}U^\dag]
    Tr[\sigma_{\hat{p}_t}V\sigma_{q_t}V^\dag]
    Tr[\sigma_{\hat{p}_c}U\sigma_{q'_c}U^\dag]
    Tr[\sigma_{\hat{p}_t}V\sigma_{q'_t}V^\dag]
\end{eqnarray}
where $(\hat{p}_c,\hat{p}_t)$ is the unique pair in 
$\{0,x,y,z\}^2$ such that
\begin{equation}
    \sigma_{\hat{p}_c}[c]\sigma_{\hat{p}_t}[t] = \pm 
    CNOT[c,t](\sigma_{p_c}[c]\sigma_{p_t}[t])CNOT[c,t] 
    \mbox{       (see Table \ref{tab:hatmap})}.
    \label{mapping}
\end{equation}
Thus $G_n(p,q,q')$ is, for fixed $V$, a bilinear function 
of $U\sigma_{q_c}U^\dag$ and $U\sigma_{q'_c}U^\dag$; and 
for fixed $U$, a bilinear function of $V\sigma_{q_t}V^\dag$ 
and $V\sigma_{q'_t}V^\dag$. Because of Requirement 
\ref{assump:rand}, we can deduce that 
$\Ex{G_n(p,q,q')\mid \psi_n,c,t}=0$ {\em unless} $q_c=q'_c$ 
and $q_t=q'_t$, i.e. $q=q'$. Moreover, in this case we have
\begin{equation}
\Ex{G_n(p,q,q)\mid \psi_n,c,t} = \left\{\begin{array}{ll} 1,&\hat{p}_c=\hat{p}_t=q_c=q_t=0;\\ \frac{1}{3},& \hat{p}_c=q_c=0, \hat{p}_t\neq 0, q_t\neq 0;\\ \frac{1}{3},& \hat{p}_c\neq 0, q_c\neq 0, \hat{p}_t=q_t=0;\\ \frac{1}{9},& \hat{p}_c\neq 0, q_c\neq 0, \hat{p}_t\neq 0, q_t\neq 0;\\ 0 & \mbox{otherwise}.
\end{array}\right.\end{equation}
It follows that
\begin{equation}\label{eq:evolutioncoef}
\Ex{\xi^2_{n+1}(p)\mid \psi_n,c,t} = \left\{\begin{array}{ll} \xi^2_n(p_{c\leftarrow 0,t\leftarrow 0}),&\hat{p}_c=\hat{p}_t=0;\\ \frac{1}{3}\sum_{w\in\{x,y,z\}}\xi^2_n(p_{c\leftarrow 0,t\leftarrow w}),& \hat{p}_c=0, \hat{p}_t\neq 0;\\ \frac{1}{3}\sum_{w\in\{x,y,z\}}\xi^2_n(p_{c\leftarrow w, t\leftarrow 0}),& \hat{p}_c\neq 0, \hat{p}_t=0;\\ \frac{1}{9}\sum_{w,w'\in\{x,y,z\}}\xi^2_n(p_{c\leftarrow w,t\leftarrow w'}),& \hat{p}_c\neq 0, \hat{p}_t\neq 0,\end{array}\right.\end{equation}
where $p_{c\leftarrow w,t\leftarrow w'}$ is the string that equals $p$ with $p_c$ and $p_t$ replaced by $w$ and $w'$, respectively. 
 Thus we may now determine a Markov Chain on the coefficients
$\xi^2(p)$.

\begin{table}[t]
 \begin{center}
 \begin{tabular*}{.333\textwidth}{|ccc|ccc|}
  \hline
 $00$ & $\leftrightarrow$ &$00$ &$x0$ & $\leftrightarrow$ & $xx$ \\
 $y0$ & $\leftrightarrow$ & $yx$ &$z0$ & $\leftrightarrow$ & $z0$ \\
 $0x$ & $\leftrightarrow$ & $0x$ &$0y$ & $\leftrightarrow$ & $zy$ \\
 $0z$ & $\leftrightarrow$ & $zz$ &$xz$ & $\leftrightarrow$ & $yy$ \\
 $yz$ & $\leftrightarrow$ & $xy$ & $zx$ & $\leftrightarrow$ & $zx$\\
 \hline
 \end{tabular*}
 \end{center}
 \caption{\label{tab:hatmap}
 The map $p_c p_t \leftrightarrow \hat{p}_c\hat{p}_t$ as defined 
 by eq. (\ref{mapping}).}
\end{table}

\subsubsection{The mapping to a Markov Chain}\label{sec:RW}
 We now notice the following key facts. First, 
$\{\xi_n^2(q)\}_{q\in \{0,x,y,z\}^N}$ is 
{\em a probability distribution} over $\{0,x,y,z\}^N$
because
$$\sum_{q}\xi_n^2(q) = Tr[\ketbra{\psi_n}^2]=1.$$

Second, consider the following way to generate a random 
$p\in \{0,x,y,z\}^N$ from an element $q=q_1\dots q_N\in \{0,x,y,z\}^N$.
\begin{enumerate}
\item choose a pair $(c,t)\in [1,\ldots,N]^2$ of distinct 
elements in $[1,\ldots,N]$, uniformly amongst all such pairs;
\item if $q_c = 0$, set $w_c=0$; else, choose $w_c\in\{x,y,z\}$ 
uniformly at random;
\item if $q_t = 0$, set $w_t=0$; else, choose $w_t\in\{x,y,z\}$ 
uniformly at random;
\item set $p=q_{c\leftarrow \hat{w}_c,t\leftarrow \hat{w}_t}$
according to the mapping eq. (\ref{mapping})
\end{enumerate}
We {\em claim} that {\em if $q$ is distributed according to $\{\xi^2_n(q)\}_q$, the distribution of $p$ is given by $\{\Ex{\xi_{n+1}(p)\mid \psi_{n}}\}_p$}. In fact, fix $c,t$. The ``hat map" $w_cw_t\leftrightarrow \hat{w}_c\hat{w}_t$ is self-inverse, hence the above choice of $p$ corresponds to
setting $\hat{p}_c\hat{p}_t=w_cw_t$ and $p_i=q_i$ for all $i\in [1,\ldots,N]\backslash\{c,t\}$. Direct inspection reveals that the probability of obtaining $p$ (given $c,t$) is given precisely by \eqnref{evolutioncoef}, which (by averaging over $c,t$) proves the claim.

We have shown that
\begin{lemma}

\label{lem:mapping}Assume that $P$ is a Markov Chain on $\{0,x,y,z\}^N$ where the transitions from a state $q\in \{0,x,y,z\}^N$ are described by the above random choices. Start this chain from $q_0$ distributed according to $\{\xi^2_0(q)\}_q$. Then the distribution $q_n$ of the state of the chain at time $n$ satisfies
$$\forall p\in\{0,x,y,z\}^N,\;\Pr{q_n=p} = \sum_{q,p}P^n(p,q)\xi_0^2(p) = \Ex{\xi_n(p)^2\mid \psi_0}.$$
As a result, if $Z_A\equiv \{p\in\{0,x,y,z\}^N\,:\, \forall i\in[1,\ldots,N]\backslash A,\, p_i=0\}$, then (by \eqnref{fundamental})
 
\begin{equation}
    \Ex{Tr[\rho_{A,n}^2]\mid \psi_0} = 2^{N_B}\Pr{q_n\in Z_A}
    = 2^{N_B} \sum_{p:\{N_n\subset A\}}\xi_n^2(p).
    \label{event}
\end{equation}

\end{lemma}

\subsection{Analysis of the Markov chain and proof of the main result}
 Now we need to analyse the Markov Chain that we have 
defined above with respect to its convergence properties.
To this end we further simplify the problem by relating it
to a simpler Markov Chain using some standard techniques
that will be outlined below.

 \subsubsection{Reduction of the chain}

 Define:
\begin{equation}\label{eq:NC}\X(q)\equiv \{i\in[1,\ldots,N]\,:\, q_i\neq 0\}.\end{equation}
One can easily show the following:
\begin{proposition}\label{prop:reduced}For all $n$, all $E\subseteq [1,\ldots,N]$, all $q$ with $\X(q_n)=E$, all $e\in E$, $d\not\in E$, if $q_0,q_1,q_2,\dots$ is an evolution of the Markov Chain $P$,

\begin{eqnarray*}\Pr{\X (q_{n+1})}
= E\cup\{d\}\mid {q_n=q}&=& 
\frac{2|E|}{3\binom{N}{2}},\\
\Pr{\X(q_{n+1}}={E\backslash\{e\}}\mid {q_n=q}&=& \frac{2(|E|-1)}{9\binom{N}{2}},\end{eqnarray*}where $|E|$ is the cardinality of set $E$.\end{proposition}

\begin{proof}Assume that we are given $q_n=q$, $\X(q_n)=E$ and consider the random procedure for $P$ described in the previous section. If $q_c=q_t=0$, then $p=q$. If $q_c=0$ but $q_t\neq 0$, then $w_c=0$ and $w_t$ is uniformly chosen from $\{x,y,z\}$; hence $p_cp_t=\hat{w}_c\hat{w}_t$ is uniform from $\{0x,zz,zy\}$; i.e. $\X(q)$ is replaced by $\X(q)\cup\{c\}$ with probability $2/3$ and remains the same with probability $1/3$. Similarly, if $q_c\neq 0$ but $q_t=0$, $t$ is added to $\X(q)$ with probability $2/3$ and stays the same otherwise. Finally, if we have $q_c\neq 0$, $q_t\neq 0$, $w_cw_t$ is chosen uniformly from $\{x,y,z\}^2$, hence $p_cp_t$ is uniform over
$$\{y0,0z,yz,xy,x0,0y,xz,yy,zx\}.$$ Thus in this case there are three possibilities: $c$ (and $c$ alone) is removed from $\X(q)$ (probability $2/9$); $t$ (and $t$ alone) is removed from $\X(q)$ (probability $2/9$); or nothing happens (probability $5/9$).
We deduce from the above that an element $d\in [1,\ldots,N]\backslash E$ can be added to $E$ only if it is one of $\{c,t\}$ and the remaining element in $\{c,t\}$ belongs to $\X(q)=E$. For each $d$ there are $2|E|$ such pairs, out of all possible $N(N-1)$; and if such a pair is chosen, $d$ is added with probability $2/3$. On the other hand, an element $e\in E$ can be removed only if it is one of $\{c,t\}$ and the remaining element is in $E$, in which case (corresponding to $2(|E|-1)$ out of $N(N-1)$ pairs), $e$ is actually removed with probability $2/9$. These assertions imply the proposition.\end{proof}

By Proposition 1, $\{\X_n\equiv \X(q_n)\}_n$ is a Markov Chain. 
Since the only event we are interested in is 
$\{q_n\in Z_A\} = \{\X_n\subset A\}$ (see eq. (\ref{event})), 
we may restrict our attention to this ``reduced" chain. For convenience, we state this as a proposition.

\begin{proposition}\label{prop:reduced2}Let $\{\X_n\}_n$ be the Markov Chain defined according to $P$ and Proposition $\ref{prop:reduced}$ above, started from a $\X_0=\X(q_0)$ with $q_0$ distributed according to the distribution $\{\xi_0^2(q)\}_{q}$ given by a state $\ket{\psi_0}$ (cf. \lemref{mapping}). Then
$$\Ex{Tr[\rho_{A,n}^2]\mid \psi_0} = 2^{N_B}\Pr{\X_n\subset A}.$$\end{proposition}
From now on, we will only deal with the ``reduced" chain $\X_n$.

\subsubsection{Dealing with the isolated state}

It should be clear that $\X_n$ as defined above is {\em not} ergodic, as the state $\emptyset$ is isolated; i.e. there are no transitions to or from it from the rest of the state space. This corresponds to the~fact that $q=0\dots 0$ is an isolated state of the initial Markov Chain.

However, this is not a problem, as we know that
$$\Pr{\X_n=\emptyset} = \Pr{\X_0=\emptyset} = \Pr{q_0=0\dots 0} = \xi^2_0(0\dots 0) = \frac{Tr[\ketbra{\psi_0}]^2}{2^N} = \frac{1}{2^N}.$$
This means that we can neglect this state and restrict our calculations with $\X_n$ to the state space $\Omega = 2^{[1,\ldots,N]}\backslash\{\emptyset\}$, as we know the contribution of $\emptyset$ to the final result.

 We now wish to show that the restricted chain is {\em ergodic}, i.e. that it has a unique stationary distribution $\sM$ for which
$$\forall E,F\in \Omega,\, \lim_{n\to +\infty}\Pr{\X_n=F\mid N_0=E} = \sM(F).$$
To prove this, it suffices \cite{AldousFill} to show that the chain is {\em irreducible} and {\em aperiodic}. Irreducibility holds when there are sequences of valid transitions between any pair of states in $\Omega$, which can be easily checked in our case. Aperiodicity means that there is no way to split $\Omega=\Omega_1\cup\Omega_2$ so that all transitions happen from a state in one of $\Omega_1$ or $\Omega_2$ to a state in the other set. But this is implied by the fact that $\Pr{\X_{n+1}=E|\X_n=E}>0$ for all $E\in\Omega$. This proves ergodicity, as desired.

\subsubsection{Stationary distribution}

We now prove that the $\X_n$ chain on $\Omega$ is {\em reversible} and determine the stationary distribution $\sM$. {\em Reversibility} means that the stationary distribution $\cM$ on $\Omega$ satisfies the {\em detailed balance condition}: for all distinct $C,D\in\Omega$,
\begin{equation}\cM(C)\Pr{\X_1 = D\mid N_0=C} = \cM(D)\Pr{\X_1 = C\mid \X_0=D}.\end{equation}
 Since we know that the chain is ergodic, the {\em existence} of a $\sM$ satisfying the above equation implies that this $\sM$ is the unique stationary distribution of the $\X_n$ process.
 
We make the ansatz that $\cM(C)=f(|C|)$ is a function of $|C|$ only. In the above comparison of $C$ and $D$, we can assume wlog that $|C|=k$ and $|D|=k+1$ for some $1\leq k\leq N-1$. Then the reversibility condition becomes
$$f(k)\frac{4k}{3N(N-1)} = f(k+1)\frac{4k}{9N(N-1)}\Rightarrow f(k+1)= 3f(k)\Rightarrow f(k) = \frac{3^k}{Z},$$
where $Z$ is a normalizing factor determined by the condition $\sum_{C\in \Omega}\cM(C)=1$, i.e.
$$Z = \sum_{C\in \Omega} 3^{|C|} = \sum_{k=1}^{N}\binom{N}{k} 3^{k} = 4^N-1.$$
Thus
$$\cM(C) = \frac{3^{|C|}}{4^N-1},\,\, C\in \Omega$$
is the unique stationary distribution of the $\{\X_n\}$ chain restricted to $\Omega$.

\subsubsection{Limits}\label{sec:limits}

Recall from \lemref{mapping} that the quantity we wish to 
estimate is $\Ex{Tr(\rho_{A,n}^2)\mid \psi_0}=2^{N_B}\Pr{\X_n\subset A}$. Using ergodicity of the chain restricted to $\Omega$, we 
know that this quantity converges as $n\to +\infty$ to
\begin{eqnarray*}
\lim_{n\to +\infty}2^{N_B}\Pr{\X_n\subset A}&=& 
2^{N_B}\left\{\frac{1}{2^N}+ \left(\frac{2^N-1}{2^N}\right)
\sum_{\emptyset\neq C\subset A}\sM(C)\right\}\\ 
&=&2^{N_B-N}\left\{1+\frac{2^N-1}{4^N-1}
\left(\sum_{\emptyset\neq C\subset A}3^{|C|}\right)\right\}\\
&=& 2^{N_B-N} \left(1 + \frac{1}{2^N+1}\sum_{k=1}^{N_A}\binom{N_A}{k}3^k\right)\\
&=& 2^{-N_A}\left(1 + \frac{4^{N_A}-1}{2^N+1}\right)\\ 
&=& 2^{-N_A}\left(\frac{2^N+4^{N_A}}{2^N+1}\right)= \frac{2^{N_B}+2^{N_A}}{2^N+1}.\end{eqnarray*}
Thus (again using \lemref{mapping})
$$\forall\psi_0,\;\lim_{n\to +\infty}\Ex{Tr(\rho_{A,n}^2)\mid \psi_0} = \frac{2^{N_B}+2^{N_A}}{2^N+1}.$$
This result can also be deduced directly from the convergence 
of $\ket{\psi_n}$ to a uniformly random state as $n\to +\infty$, 
together with known formulae for the expected purity.

\subsubsection{Mixing of the reduced Markov Chain}

Our main goal in this section is to prove bounds on the mixing time of the Markov Chain given by $\X_n$. We will take an indirect route to do so.

\begin{lemma}\label{lem:gap}The Markov chain given by $\X_n$ has spectral gap $\geq 4/9N(N-1)$.\end{lemma}
\begin{proof}Consider a chain $B_n$ on $\Omega$ defined as follows. Assume $B_n=B$ and choose a $1\leq j\leq N$ uniformly at random. If $j\in B$ and $|B|\geq 1$, set $B_{n+1}=B\backslash \{j\}$ with probability $1/3$ and $B_{n+1}=B$ with probability $2/3$. If $j\not\in B$, set $B_{n+1}=B\cup\{j\}$. We {\em claim} that
\begin{claim}\label{claim:Bn}$B_n$ is reversible, ergodic and has the same stationary distribution $\sM$ as $\X_n$. Moreover, the spectral gap of $B_n$ is at least $1/3N$.\end{claim}
Before proving the claim, we show how it implies the lemma, i.e. the $\geq 4/9N(N-1)$ bound for the spectral gap of $\X_n$. This is possible via a {\em comparison} of Markov chains. By Theorem 2.14 in \cite{MontenegroTetali}, it suffices to show that for all distinct $C,D\in\Omega$
\begin{equation}\label{eq:comparison}\sM(C)\Pr{\X_1=D\mid \X_0=C}\geq \frac{4\sM(C)}{3(N-1)}\Pr{B_1=D\mid B_0=C}.\end{equation}
Indeed, because both $\X_n$ and $B_n$ are reversible chains, both the LHS and RHS are symmetric in $C,D$. Therefore, in proving eq. 
\eqnref{comparison} we can assume that $D=C\cup\{j\}$ for 
some $j\in[1,\ldots,N]\backslash C$. Then it is easy to see that
\begin{eqnarray*}\sM(C)\Pr{\X_1=D\mid \X_0=C} = 
\frac{2|C|\sM(C)}{3\binom{N}{2}} \geq \frac{4\sM(C)}{3(N-1)}\Pr{B_1=D\mid B_0=C}.\end{eqnarray*}
To finish, we must prove Claim $\ref{claim:Bn}$.

\begin{proof}{\em (of Claim $\ref{claim:Bn}$)}Reversibility of $B_n$ follows from the fact that if $C\in \Omega$, $D=C\cup\{d\}\in \Omega$ with $d\not\in C$ are given,
$$\sM(C)\Pr{B_1=D\mid B_0=C} = \frac{\sM(C)}{N} = \frac{\sM(D)}{3N} = \sM(D)\Pr{B_1=C\mid B_0=D}.$$
We use the {\em path-coupling} technique of Bubler and Dyer (see e.g. \cite{AldousFill}) to prove ergodicity and the desired spectral gap bound for $B_n$. For $B,B'\in\Omega$, let $d(B,B') = |B\Delta B'|$ be the Hamming distance between $B$ and $B'$. Call $B$ and $B'$ {\em adjacent} if $d(B,B')=1$. We will show that one can couple one-step evolutions $B_1$, $B'_1$ started from adjacent $B_0$, $B'_0$ so that
\begin{equation}\label{eq:pathcoupling}\Ex{d(B_1,B'_1)-d(B_0,B'_0)} = -\frac{1}{3N}.\end{equation}
This means (cf. reference \cite{AldousFill}) that for {\em arbitrary} $B_0$, $B'_0$, there is a coupling of $B_1,\dots,B_n$ and $B'_1, \dots, B'_n$ such that for all $n\geq 0$,
\begin{equation}\label{eq:extcoupling}\Ex{d(B_{n},B'_{n})}\leq \left(1-\frac{1}{3N}\right)^nd(B_0,B'_0)\leq N\left(1-\frac{2}{3N}\right)^n.\end{equation}
From this it follows via a standard argument that the 
statistical distance/ between $B_n$ and $B_n'$ decays 
at an exponential rate of $\leq (1-1/3N)$, which also implies the desired spectral gap bound.

The coupling in \eqnref{pathcoupling} is indeed very simple. Suppose we are given adjacent $B_0$ and $B'_0$. Without loss of generality, we can assume that $B_0=\{1,\dots,k\}$ and $B'_0=\{1,\dots,k+1\}$ for some $1\leq k\leq N-1$. We choose a $j\in[1,\ldots,N]$ uniformly at random and update $B_0$ and $B'_0$ in the following way:
\begin{enumerate}
\item If $1\leq j\leq k$,
\begin{enumerate}
\item if $k\geq 2$, set $B'_1 = B'_0\backslash \{j\}$ and $B_1 = B_0\backslash \{j\}$ with probability $1/3$ and do not change the states with probability $2/3$;
\item else if $k=1$ do this only for $B'_1$, leaving $B_1=B_0$ always.
\end{enumerate}
\item else if $j=k+1$,
\begin{enumerate}
\item set $B_1=B_1'=B'_0$ with probability $2/3$
\item OR $B_1=B'_0$, $B'_1=B_0$ with probability $1/3$;
\end{enumerate}
\item else if $i+2\leq j\leq N$, set $B_1=B_0\cup\{j\}$, $B_1'=B'_0\cup \{j\}$.
\end{enumerate}
Clearly, $B_1$ and $B'_1$ each have the right distribution. Moreover, the distance between them increases by one (with probability $1/3$) in case $1.b$, decreases by $1$ in case $2.a$, and remains the same in all other cases. It follows that
\begin{eqnarray*}\Ex{d(B_1,B'_1)-d(B_0,B'_0)} &=& -\frac{2}{3N} + {\Ex{1_{\{k=1,1\leq j\leq k\}}}}\\
&\leq & \frac{1-2}{3N} = -\frac{1}{3N},\end{eqnarray*}
i.e. eqns. \eqnref{pathcoupling} and \eqnref{extcoupling} follow.\end{proof}

{\em (End of proof of \lemref{gap}.)}\end{proof}

\subsubsection{End of proof of the main theorem}

We have shown in the Introduction that the Theorem follows 
from \lemref{main}. Moreover, Proposition \ref{prop:reduced2} 
shows that (omitting the initial state),
$$\Ex{Tr(\rho_{A,n}^2)}=2^{N_B}\Pr{\X_n\subset A},$$
where $N_0$ has distribution given by $\ket{\psi_0}$ 
in the manner described above. Notice that
$$\Pr{\X_n\subset A} = \Pr{\X_0=\emptyset} + \Pr{\X_0\neq \emptyset,\X_n\subset A} = \frac{1+(2^N-1)\Pr{\X_n\subset A\mid \X_n\neq \emptyset}}{2^N}.$$
Using the ergodicity of $\X_n$ restricted to $\Omega$ and the limit formula in Section $\ref{sec:limits}$,
$$\left|2^{N_B}\Pr{\X_n\subset A} - \frac{2^{N_A}+2^{N_B}}{2^N+1}\right|\leq 2^{N_B}\|\X_n - \sM\|_{sd},$$
where $\|\X_n-\sM\|_{sd}$ is the 
statistical distance between $\sM$ and the distribution 
of $\X_n$ restricted to $\Omega$. However, we know that
\begin{enumerate}
\item the restriction of $\X_n$ is ergodic and reversible;
\item its spectral gap is bounded below by $4/9N(N-1)$;
\item for any $E\in \Omega$, the probability that $\X_{n+1}=E$ given $\X_{n}=E$ is (cf. Proposition $\ref{prop:reduced}$)
\begin{eqnarray*}\Pr{\X_{n+1}=E\mid \X_{n}=E} &=&1 - \sum_{e\in [1,\ldots,N]\backslash E}\frac{2|E|}{3\binom{N}{2}}-\sum_{e\in E}\frac{2(|E|-1)}{9\binom{N}{2}}\\
&=& 1 - \frac{2|E|(N-|E|)}{3\binom{N}{2}} - \frac{2|E|(|E|-1)}{9\binom{N}{2}}\\
&=& 1 - \frac{2\left((N-1)|E|-\frac{4}{3}|E|^2\right)}{3\binom{N}{2}}.\end{eqnarray*}
A simple calculation shows that the numerator above is at most $3(N-1)^2/8\leq 3/4\binom{N}{2}$ and that $\Pr{\X_{n+1}=E\mid \X_{n}=E}\geq 1/4$ always.
\end{enumerate}
It follows from standard Markov chain theory (eg. Corollary 1.15 in \cite{MontenegroTetali}) that all eigenvalues of the $\X_n$ chain that are different from $1$ lie between $-3/4+1/4 =-1/2$ and $1-4/9N$ and that for any initial distribution of $\X_0$
$$\|\X_n-\sM\|_{sd}\leq \frac{\left(1-\frac{4}{9N(N-1)}\right)^n}{\sqrt{\min_{C\in\Omega} \sM(C)}}\leq 2^N\exp\left(-\frac{4n}{9N(N-1)}\right).$$
Since $2^{N_B}\leq 2^N$, this implies that
$$\left|\Ex{Tr(\rho_{A,n}^2)} - \frac{2^{N_A}+2^{N_B}}{2^N+1}\right|\leq 4^{N}\exp\left(-\frac{4n}{9N(N-1)}\right),$$
which finishes the proof of Theorem 1.

\section{Efficient simulation of process on classical computer}
\label{sec:stabilizers}
We now show that the purity evolution during this random 
process, which corresponds to a randomly chosen quantum 
computer circuit, can be simulated efficiently on a classical 
computer in the sense that the same statistics can be 
obtained with a polynomial effort in $N$. The efficient 
simulation is achieved by noting that one can use stabilizer 
states, a discrete and finite subset of general quantum 
states which can be parameterised efficiently (for a brief
introduction to stabilizer states see Appendix C). This 
is of interest 
since it may lead to methods to efficiently simulate 
quantum systems with a high degree of entanglement.

Stabilizer states, despite having various restrictions, 
possess a rich entanglement structure exhibiting multi-partite 
entanglement 
\cite{gottesman1,gottesman2,DP,Audenaert P 05,fattal,Hein EB 04}
and may be used in the approximate description of ground states
of Hamiltonians \cite{AndersPDVB06}. 
Here we will use two general facts about stabilizer states. 
Firstly we note that for a single qubit there are six stabilizer 
states, given by the +1 eigenvectors of the Pauli operators 
$\sigma_1,-\sigma_1,\sigma_2,-\sigma_2,\sigma_3,-\sigma_3$. One can note that these are evenly distributed 
over the Bloch Sphere, so are similar in that regard to the 
unitarily invariant measure on U(2). Secondly we use the known 
fact that \{H,S,CNOT\} form a universal set for stabilizer gates, 
in that any gate that maps the set of stabilizer states onto 
and into itself can be decomposed as a combination of those.  

The entanglement probability distribution on stabilizer states 
is derived in \cite{DP}. In a system of $N$ spins where 
$N_A$($N_B$) is Alice's (Bob's) number of qubits the probability 
of finding that entanglement between $A$ and $B$ in a randomly 
chosen pure stabilizer state equals $E$ is given by
\begin{eqnarray}
P(E)=\frac{\prod_{i=1}^{N_A}(2^i\!+\!1)}{\prod_{k=N-N_A\!+\!1}^{N}(2^k\!+\!1)}
\!\prod_{j=1}^{E}\!\!
\frac{ \left(2^{N\!-\!N_A\!+\!1\!-\!j}\!-\!1\right)\!\!\left( 2^{N_A\!+\!j}\!-\! 2^{2j-1}\right)}
{2^{2j}-1 }
\label{eq:stabprob}
\end{eqnarray}
where $E$ is an integer. This can be used to show the average 
is nearly maximal and the distribution squeezes up around the 
average with increasing N.

Considering the random walk at hand, we note that the 
expected purity of a flat distribution of stabilizer 
states is the same as that for the unitarily invariant 
measure on general states\cite{divincenzo}.

So the stabilizer random walk described earlier will asymptotically 
yield the same expected purity as the general state walk\footnote{An alternative way of explaining why the asymptotic time average of the purity
is the same, whether using the stabilizer circuit or the general state circuit, is the concept of 2-designs; see \cite{dankert,gross}}.

Each one-qubit stabilizer gate permutes the six states. The stabilizer gate invariant distribution is therefore 
that which is flat on these states.

\begin{lemma}\label{lem:stabilizers}Assume that $T$ is drawn from a flat
distribution of the 6! single qubit stabilizer gates.
Then requirement 1 is satisfied.\end{lemma}

\begin{proof}
In the notation defined in Lemma \ref{lem:sigmas} in appendix B, the 
six stabilizer states are given by $(r_x,r_y,r_z)=
(1,0,0), (-1,0,0), (0,1,0), (0,-1,0), (0,0,1), (0,0,-1)$.
One can use the proof that Haar measure on U(2)
satisfies requirement 1 which is provided in Appendix B, but replacing the Haar measure on U(2) with the flat distribution on the six stabilizer states. 
\end{proof}

Requirement 1 is is the 
only requirement made about how we pick the single qubit gates 
U and V in the proof of Theorem 1. Accordingly, with regard to the evolution of the expected purity, the stabilizer random walk is equivalent 
 on average to that on general states when concerned
with the average purity of subsystems.

\begin{figure}
 \begin{center}
  \begin{minipage}[t]{0.5\linewidth}
   \includegraphics[width=\linewidth]{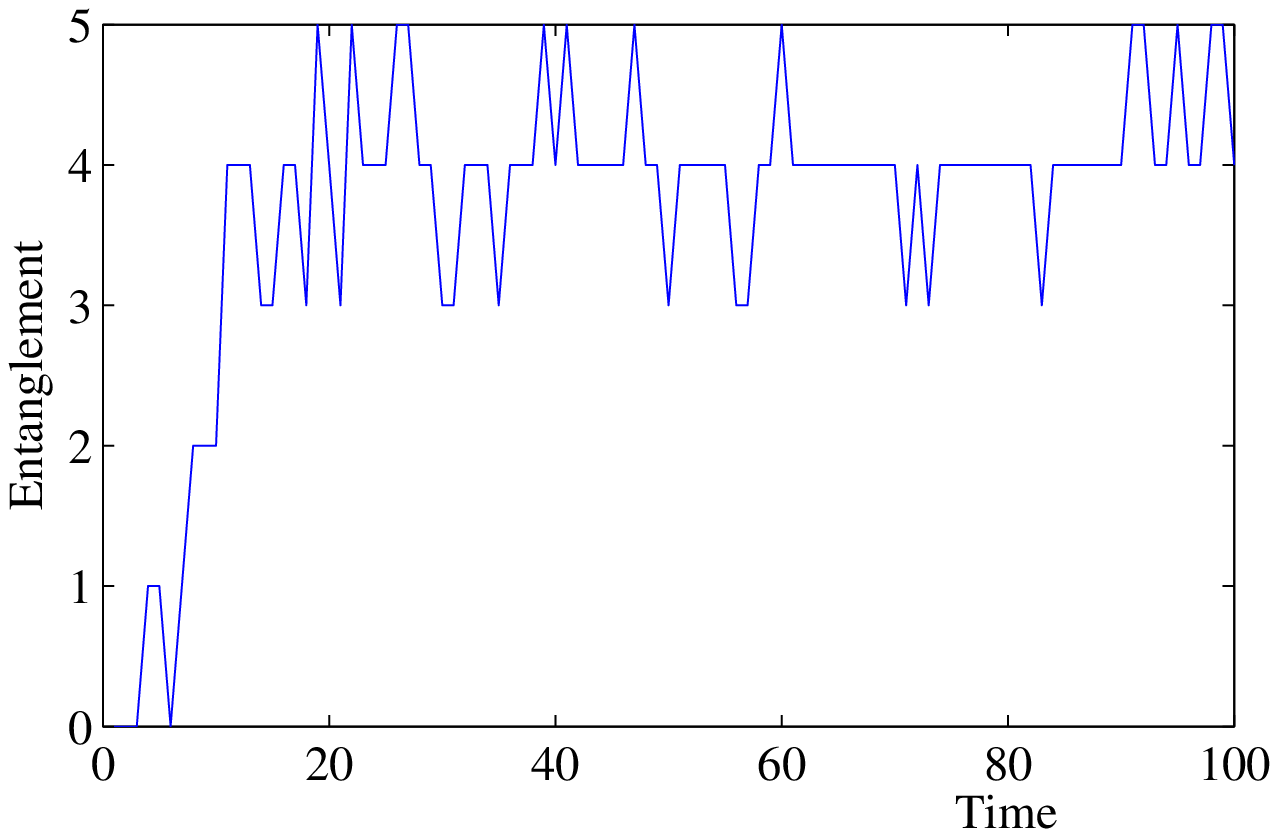}
   \caption{Entanglement evolution of stabilizer states where $N=10$ and $N_A=5$. Here a single realization only. The dramatic jumps
 reflect the fact that stabilizer entanglement only comes in integer values.\label{fig:S10A1to5i}}
  \end{minipage}\hfill
  \begin{minipage}[t]{0.5\linewidth}
    \includegraphics[width=\linewidth]{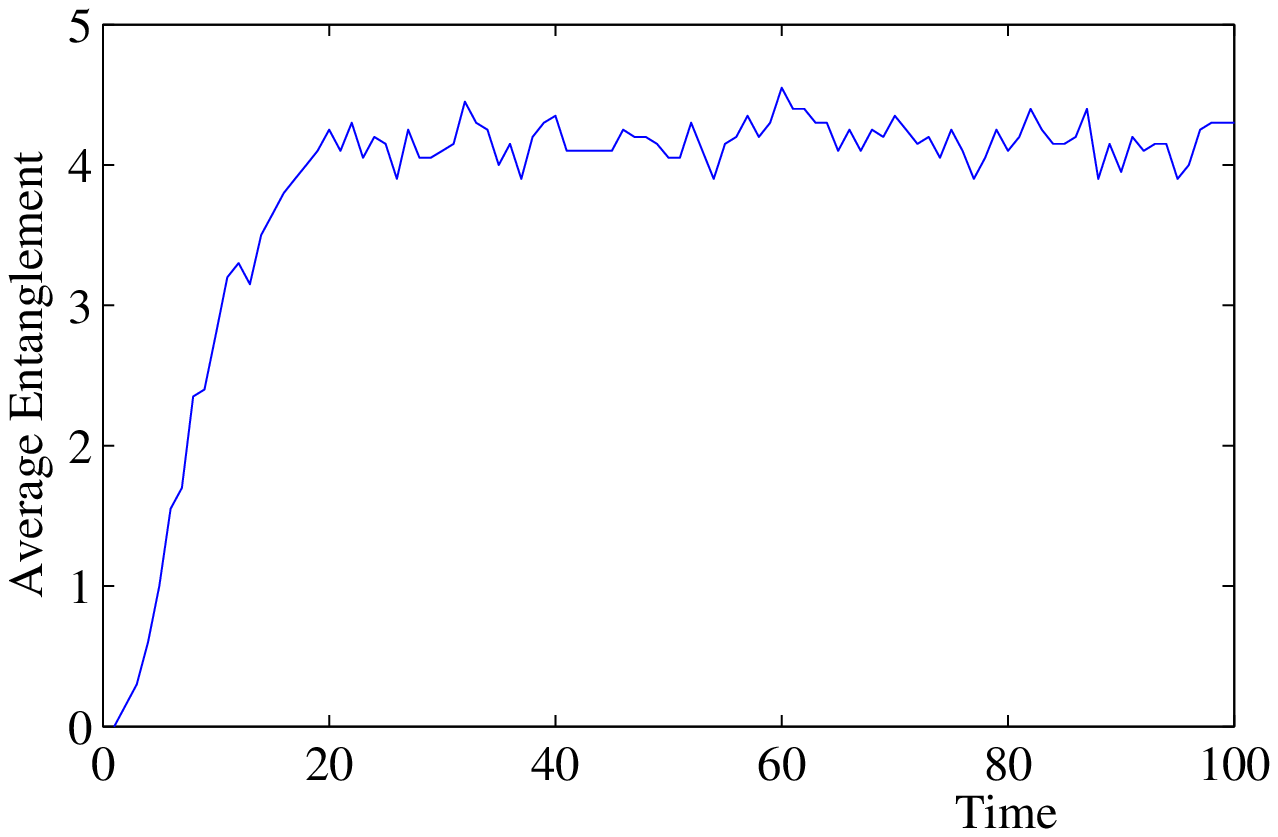}
   \caption{Here averaged over 20 realizations. One sees that the average behaviour is like that of general states\label{fig:S10A1to5ii}.}
  
  \end{minipage}
 \end{center}
\end{figure}

Our result explains why averaging the entanglement 
evolutions of different stabilizer random walk 
realizations reproduces the behaviour of the corresponding 
general state two-party process. An example of this can 
be seen in the numerical simulations shown in figures \ref{fig:S10A1to5i} and \ref{fig:S10A1to5ii}.

\section{Two phases in approach to typical entanglement}
Theorem 1 tells us that systems undergoing this sort of 
random interactions will tend to become maximally entangled 
very fast. Numerical studies in addition reveal a finer 
structure in the approach, namely two phases, apparently 
separated by a well-defined and short transition interval. 
\begin{enumerate}
\item A phase in which entanglement is rapidly increasing and spreading through 
the system.
\item A phase in which entanglement has spread through the entire
system.
\end{enumerate}

We identify three ways of defining the moment of time 
that separates these two phases that will be 
explained in the following: the saturation moment 
$\tau_{sat}$, the cutoff moment $\tau_{cutoff}$ or 
the volume scaling moment $\tau_{vol}$. 

\subsection{Saturation moment}
\begin{figure}[th]
\centerline{
\includegraphics[width=12cm]{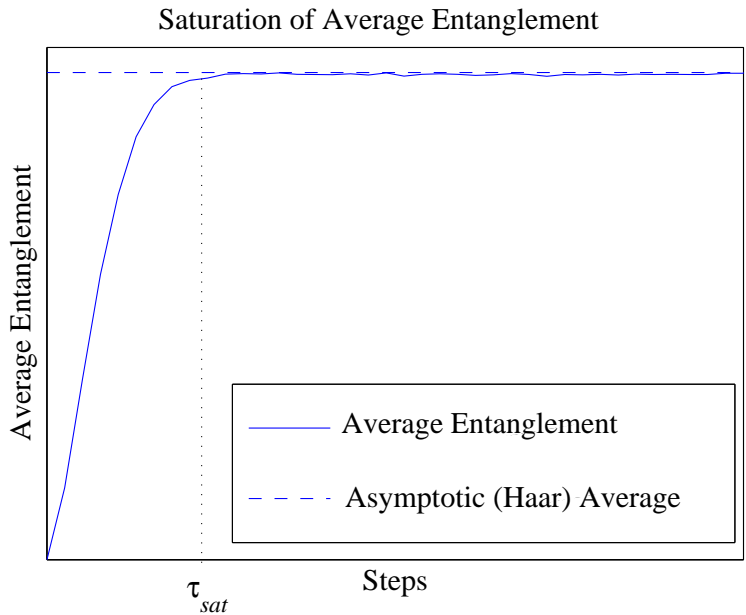}
}
\vspace{.0cm} \caption{Illustrating $\tau_{sat}$. We find that this behaviour is generic from numerical simulations.}
\label{fig:saturationtime}
\end{figure}

Applying the random interaction we make the following 
\\
{\bf Observation:} There is a moment of saturation of the average entanglement.
\\
\\
Before this moment the average block 
entanglement increases essentially linearly 
in the time $n$. After the transition moment 
it is however practically constant and nearly 
maximal. Therefore we term the transition time
between the two regimes the 'saturation time' $\tau_{sat}$. 
There is some degree of freedom in what exact value to assign to this time.
One could for example specify $\tau_{sat}(\epsilon)$ as the moment that the gradient of the average entanglement
curve is within some fixed accuracy $\epsilon$ to 1.
Figure \ref{fig:saturationtime} shows how 
this moment is reached.

\subsection{Cut off moment}

The numerical observation that there appears to be two 
time-scales involved here, first a rapid approach to 
the asymptotic value and then a slow one, led us to 
study the statistical mathematics literature for tools 
that quantify this. In fact there has been extensive 
study of such problems, motivated by the fact that a 
randomisation process, such as shuffling cards in a 
casino, is in practise performed only a finite number of times.
The question is then, how many shuffles are necessary before 
one is certain, for all practical purposes, that the cards 
are shuffled. In the setting concerned here, that 
corresponds to asking when we are certain, for practical 
purposes, that the entanglement probability distribution has achieved its asymptotic form. The tool we 
will use here is the so called "cut-off effect", which 
is exhibited by many Markov Chains\cite{diaconis}.

The cut-off refers to an abrupt approach to the stationary 
distribution occurring at a certain number of steps taken 
in the chain. Say we have a Markov chain defined by its 
transition matrix P, and that it converges to a stationary 
distribution $\pi$. Initially the total variation distance 
$TV=\parallel P-\pi \parallel=sup \mid P(E)-\pi(E) \mid $ 
between the corresponding probability distributions is 
given by $TV=1$. After $k$ steps this distance is given by
$TV(k)= \parallel P^k-\pi \parallel$. A cut-off occurs, 
basically, if $TV(k) \simeq 1$ for $k=0,1,2,...a$ and 
thereafter falls quickly such that after a few steps 
$TV(k) \simeq 0$. As we increase the size of the state 
space, the number of steps during which the abrupt approach 
takes place should decrease compared to $a$, the number 
of steps necessary to reach the cut off. Then, for very large 
state spaces, we can say that the randomisation occurs at 
$a$ steps, some function of the size of the state space. 

\begin{figure}[th]
\centerline{
\includegraphics[width=12cm]{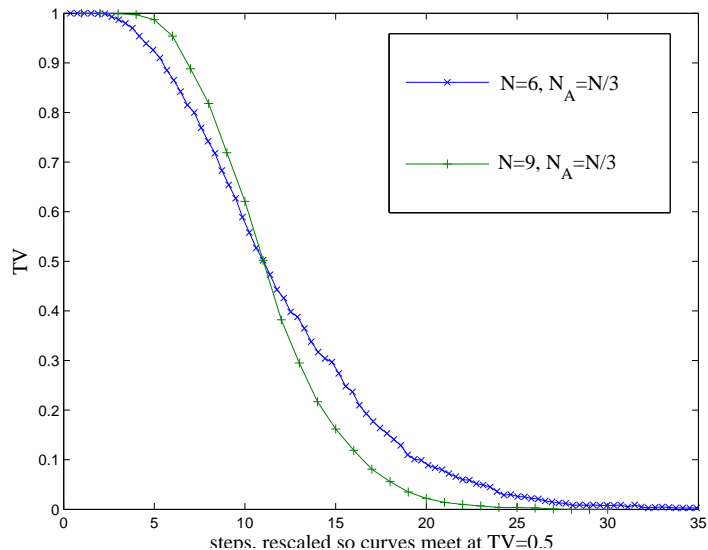}
}

\vspace{.0cm} \caption{Observe cut-off effect for general states. The total variation distance to the asymptotic entanglement probability distribution, $TV\simeq 1$ for some finite time, before falling to 0. The fall becomes more dramatic with increasing system size. 
Note that, as is the customary way of representing this effect, we have rescaled the curves to meet where $TV \simeq 0.5$. }
\label{fig:generalcutoff}
\end{figure}

As a precise definition of a cut off we use \cite{diaconis}.\\
 {\bf Definition:} \emph{
Let $P_n$, $\pi_n$ be Markov Chains on sets $\chi _n$. Let 
$a_n$, $b_n$ be functions tending to infinity, with 
$\frac{b_n}{a_n}$ tending to zero. Say the chains satisfy 
an $a_n$, $b_n$ cutoff if for some starting states $x_n$ and 
all fixed real $\theta$ with $k_n=\lfloor a_n+\theta b_n \rfloor$, 
then 
\begin{equation}
\parallel P_n^{k_n}-\pi_n \parallel\longrightarrow c(\theta)
\end{equation}
with $c(\theta)$ a function tending to zero for $\theta$ 
tending to infinity and to 1 for theta tending to minus infinity.}
\\
\\
With this definition it is now possible to study whether there
is a sharp cut-off time associated with the entanglement probability distribution, a functional on our specific Markov 
process. Indeed, numerically we make the \\
{\bf Observation:} We observe an apparent cut off effect 
in the entanglement probability distribution under the 
two-particle interaction random process described in
this work.
\\
\\
Figure \ref{fig:generalcutoff} shows an example of a cut off for general states. The state 
space has been discretised by rounding off entanglement values 
to the nearest integer. We observe that $TV\simeq 1$ for a 
while and then falls. Finally there is a stage where $TV\simeq 0$. 
The effect becomes more pronounced with increasing $N$.

\begin{figure}[th]
\centerline{
\includegraphics[width=12cm]{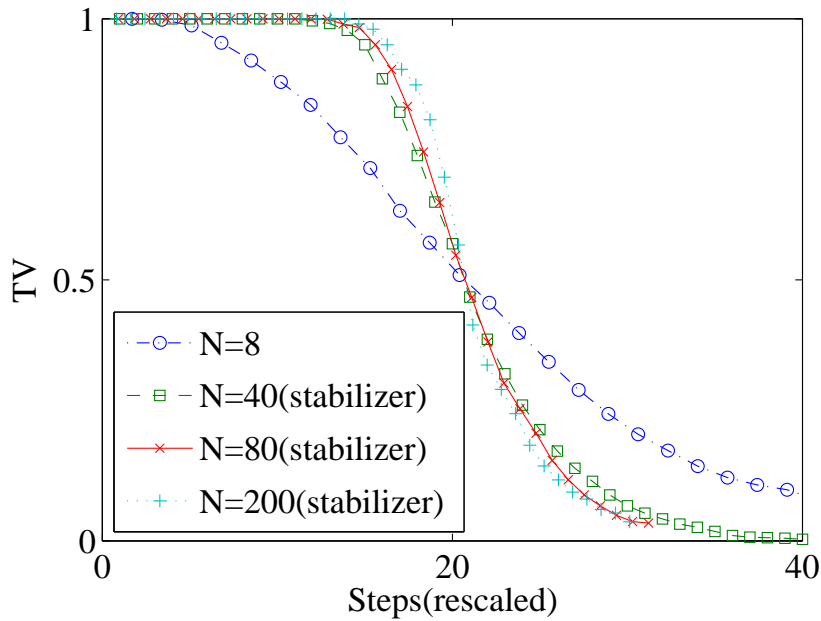}
}
\vspace{.0cm} \caption{The cut-off effect in the total variation distance ($TV$) is here studied for larger $N$ by employing stabilizer states
and tools to efficiently evaluate their entanglement\cite{Audenaert P 05}. One sees that the effect becomes increasingly pronounced with increasing $N$.
We conjecture that the function is a step function in the limit of $N \rightarrow \infty$, consistent with the normal behaviour of the cut-off effect.}
\label{fig:cutoff}
\end{figure}

Now consider the analogous situation for stabilizer states. 
From Lemma \ref{lem:stabilizers} we would expect this behaviour 
to be representative of that of general states. Since stabilizer 
states are efficiently parameterised, using them this will allow us 
to scale further with N.
\\
\\
{\bf Observation:} We observe an apparent cut off effect in 
the entanglement probability distribution under the stabilizer 
two-particle interaction random process described in this 
work.
\\
\\
How this squeezes up is showed for individual runs, averaged over 1000 realizations, in figure \ref{fig:cutoff}.
\\

As stated, the cut off effect is common in classical 
Markov Chains such as card shuffling \cite{diaconis}. 
It is a testament to the universality of mathematics 
that applying quantum gates at random to a quantum 
register apparently exhibits the same features, in 
this regard, as applying shuffles to a deck of cards. 

We term this the cut off moment $\tau_{cutoff}$. Using 
this moment to separate the first and second phase 
has the advantages that it unambiguously gives one moment, 
and this moment corresponds to the point when the entanglement 
distribution equals, for practical purposes, the asymptotic 
one.

\subsection{The area to volume scaling transition}
\begin{figure}[th]
\centerline{
\includegraphics[width=7cm, height=7cm]{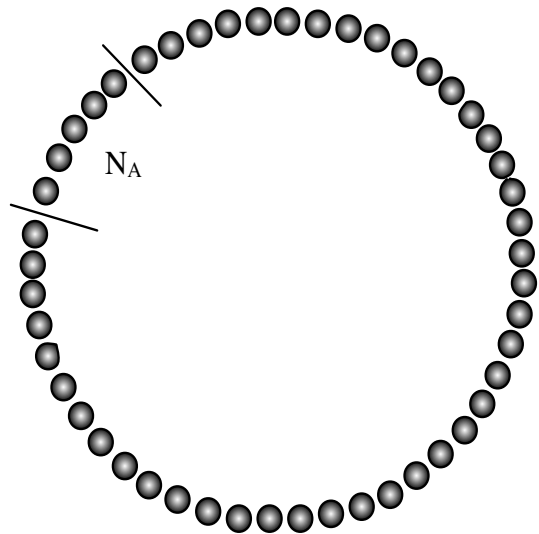}
}
\vspace{.0cm}
\caption{$N$ qubits with nearest neighbour interactions. 
We vary $N_A$ at different times and ask what the typical 
entanglement is.}
\label{fig:qubitring}
\end{figure}

\begin{figure}[th]
\centerline{
\includegraphics[width=\linewidth]{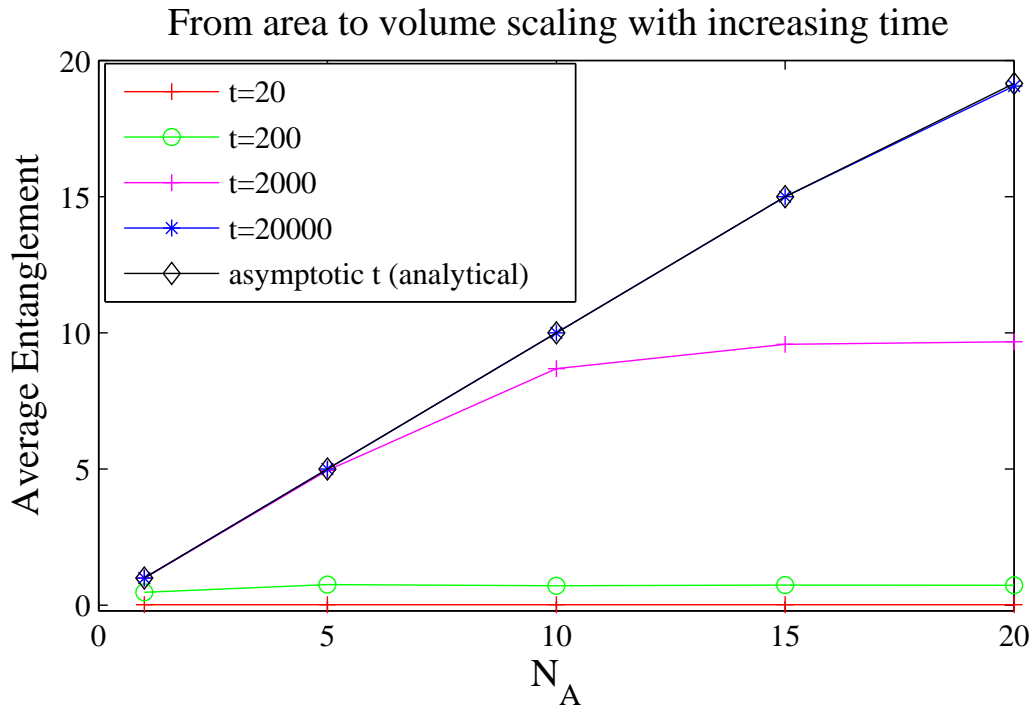}
}
\vspace{.0cm}
\caption{$N=40$ qubits with nearest neighbour interactions. 
We vary $N_A$ at different times and ask what the typical 
entanglement is. Sufficiently high statistics were taken 
that the statistical error was not visible on this plot. The asymptotic curve was calculated using equation \ref{eq:stabprob}.}
\label{fig:AVN40manytimes}
\end{figure}

We now consider the two-party random process restricted 
to nearest-neighbours interactions, as in figure \ref{fig:qubitring}.
This helps us to relate two pictures of typical entanglement. 
One is the notion that following a randomising process 
entanglement appears all pervasive in the system. In this
 case one may expect that the entanglement between 
two blocks scales roughly as the size of the smaller part, 
i.e. we observe a volume scaling. While this is the behaviour
for generic states it is in strong contrast to the behaviour
of pure states that are typically appearing in nature such as 
the ground states of Hamilton operators describing systems
with short range interactions. Indeed, in such systems the
block entanglement in ground states has been proven to scale 
as the boundary surface area between the two blocks for a 
wide variety of systems \cite{AudenaertEPW02,PlenioEDC05,CramerEPD06,KeatingM04}.
This indicates that ground states of short-range Hamiltonians tend
to explore only a tiny fraction of the entire Hilbert space
as their entanglement properties are indeed rather atypical.
\\
\\
{\bf Numerical Observation:} The entanglement scales as the 
area at small times and as the volume of the smaller
block at large times.
\\
\\
This observation motivates the introduction of the 
transition time from area to volume scaling $\tau_{vol}$
and is shown in figure \ref{fig:AVN40manytimes}.
The result could have been expected since the nearest 
neighbour structure is only respected at small times 
since at longer times the two-qubit unitaries have 
combined to form global unitaries. To assign an exact value to this moment one can specify $\tau_{vol}(\epsilon)$ 
as the moment $max |\Ex{S(\rho_A)}-\Ex{S(\rho_{A,\tau})}|\leq \epsilon $ where 
the maximisation is over all partitions.

\section{Multipartite and mixed states}
So far we have considered pure bipartite states. It is 
known that the unitarily invariant measure is not uniquely 
defined for mixed states however. This can be visualised 
already in the case of one qubit. On one qubit a unitary 
time evolution corresponds to a rotation of the Bloch Sphere. 
 Pure states lie on the surface of the sphere and the 
unitarily invariant measure is an uniform distribution
on the surface of the sphere. However 
mixed states can lie anywhere in the ball inside the
 sphere and there is no longer a unique measure as
unitary invariance does not provide any restrictions
on the radial distribution. Mixed measures may be induced 
by considering environments that allow for purifications
of mixed states. Then the measure will be obtained by using the
uniform measure on the purifications and then determine the
measure that is obtain by partial tracing. Nevertheless
some ambiguity remains related to the size of the environment.
As is evident from the results here, increasing the size 
of the environment concentrates the measure more and more 
around the maximally mixed state.

\begin{figure}[th]
\centerline{
\includegraphics[width=12cm]{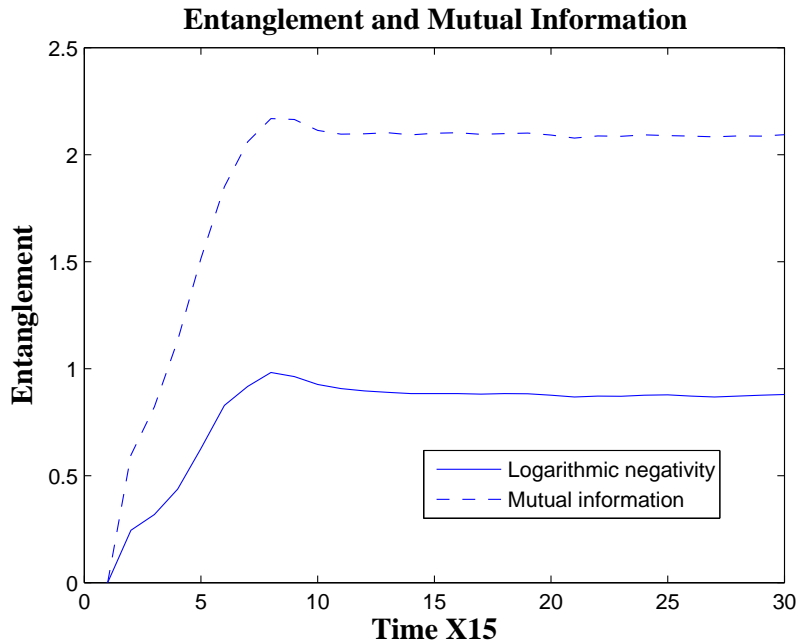}
}
\vspace{.0cm}\caption{An example of a numerical simulation showing that the average classical and quantum correlations behave similarly.
The quantum correlations are quantified using the logarithmic negativity. The combination of classical and quantum correlations is quantified using the mutual information. 
Here $N=10$, $N_A=3$, $N_B=3$ and $N_C=4$.}
\label{fig:A2to4iinew}
\end{figure}

Before evaluating the entanglement between A and B one 
traces out the environment C. This leads to AB to be 
in a mixed state. In fact, with increasing size
of the environment the state of AB tends to become more
mixed and the entanglement between A and B becomes 
negligible and ultimately disappears. This is a possible 
argument for emerging classicality. To measure entanglement 
in the mixed state we used the logarithmic negativity, $E_N$, 
defined in \cite{logneg,Martinlogneg,Eisert} and 
proven to be an entanglement monotone in \cite{Martinlogneg}. 
\begin{equation}
    E_N(\rho)=log_2 ||\rho ^{T_A}||_1.
\end{equation}
Hence here the entanglement is quantified as $E_N(tr_{C}(\rho_{ABC}))$
One can also note that for mixed states there are classical 
correlations. The mutual information, $S(\rho_A)+S(\rho_B)-S(\rho_{AB})$,
can be interpreted as the combination of classical and quantum 
correlations \cite{groisman}. \\
\\
{\bf Observation:}
The mutual information behaves very similarly to the quantum correlations (entanglement)in the two particle-interaction random walk simulations. 
\\
\\
This suggest the classical correlations
behave similarly to the entanglement in the random two-party process, as in figure \ref{fig:A2to4iinew}.
 
\begin{figure}[th]
\centerline{
\includegraphics[width=12cm]{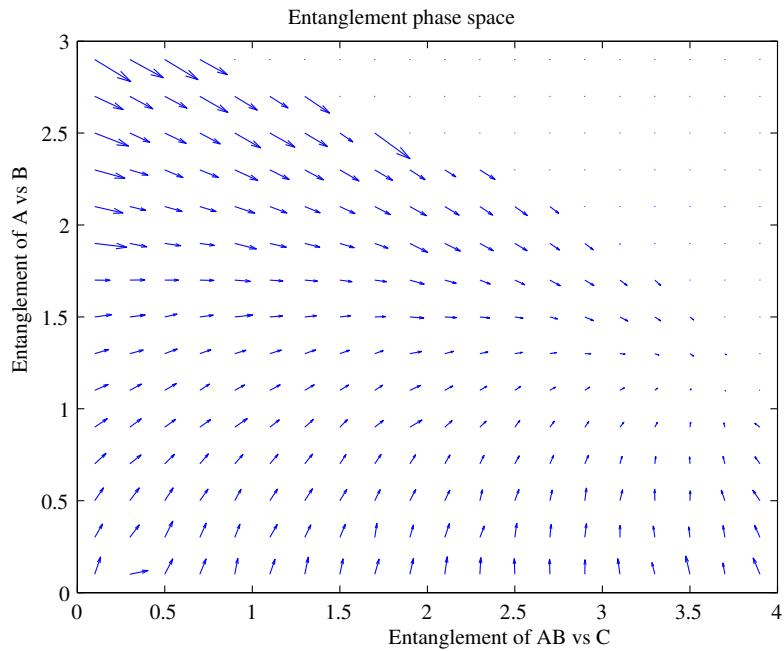}
}
\vspace{.0cm}
\caption{Entanglement 'Phase space' flow diagram. The arrows
point in the average flow direction and the length indicates the speed. The entanglement between A and B is quantified as $E_N(tr_{C}(\rho_{ABC}))$ and 
the entanglement between AB and C using $S(\rho_C)$ We see how different initial states will all tend to the attractor point. The
horizontal coordinate of this point is given by equation \ref{eq:page}. The vertical coordinate
by $\Ex{E_N(tr_{C}(\rho_{ABC}))}$. 
Here $N=10$, $N_A=3$, $N_B=3$ and $N_{C}=4$. The region in the 
top right corner is empty as such states
are not physically realizable.}
\label{fig:vphasespace100000}
\end{figure}

The usefulness of the entanglement as a possible macroscopic 
parameter is highlighted in figure \ref{fig:vphasespace100000} 
which shows how the equilibrium values is an attractor point 
regardless of initial state and that there is a flow towards 
this point. The latter observation hints that the average 
entanglement is a good parameter also in the approach to the 
attracting equilibrium. 

\subsection{Simplified multipartite description}
Considering generic entanglement only will hopefully 
provide simplifications in various settings. Here we 
suggest a way in which it simplifies the multipartite
setting. We consider the tripartite setting. Let $N$ 
qubits be shared between three parties of size $N_A$, 
$N_B$ and $N_C$, s.t. $N_A+N_B+N_C=N$ and $N_A\leq N_B\leq N_C$. 
We can then consider many types of bipartite cuts. 
Let $E(A|B)$ signify the (unitarily invariant) 
entanglement average of the entanglement across the 
$A|B$ cut after C has been traced out, and let $E(A+B|C)$ 
signify the corresponding quantity across the $A+B\|C$ cut. 
The set $\Omega$ of all possible bipartite entanglement 
averages is then 
$\Omega=\{E(A|B),E(A|C),E(B|C),E(A+B|C),E(A+C|B),E(B+C|A)\}$.
Let these values specify 'the tripartite entanglement'. 
Now for an arbitrary state, it is not the case that $\Omega$ is uniquely determined by $E(A|B)$ together with $E(A+B|C)$.
However we make the following claim. 

\emph{Conjecture}: $\Omega$ is uniquely determined by $E(A|B)$ 
together with $E(A+B|C)$

\emph{Support for conjecture:}
Firstly note that if $N_A$, $N_B$ and $N_C$ are specified then $\Omega$ 
is uniquely determined. Then we need the following two 
statements to be true to prove the claim:\\

1.$N_A+N_B$ and $N_C$ are uniquely specified by $E(A+B|C)$.
This is true if the Von Neumann entropy of entanglement 
increases monotonically when a qubit is given from the 
larger of the two parties to the smaller. We are concerned with the limit of large $N$,
where equation \ref{eq:page} implies that
$N_C=\lceil E(A+B|C) \rceil$ assuming $N_C \leq N_A+N_B$\\
 
2. For the given $N_C$ and $N$, $N_A$ and $N_B$ are uniquely 
determined by $E(A|B)$. Again one should be able to use a 
monotonicity argument here, although it will be more difficult 
due to the lack of a neat closed form of 
$E(A|B)=\Ex{E_N(tr_{C}(\rho_{ABC}))}$. It seems reasonable 
to expect it to be monotonous though since since $N_C$ is fixed.\\

This idea could presumably be extended to more than 
three parties and may lead to a description that is 
highly useful by virtue of requiring very few parameters. 

\section{Discussion and Conclusion}
The entanglement evolution during random two-qubit 
interactions was studied in some depth and we have 
provided a detailed proof of the result presented
in \cite{oliveira} that the average entanglement 
approaches the unitarily invariant value within $O(N^3)$ 
steps. The essential idea of the proof is a map from the
evolution in state space onto one on a much smaller 
space that tracks the evolution of the purity of the reduced 
subsystem. We then proved that the process can be 
simulated efficiently on a classical computer using 
stabilizer states. 

We found through numerical studies that there are two 
phases in the approach: first a phase during which 
entanglement is rapidly spreading through the system, 
and then a phase where the entanglement is suffused 
throughout the system. Three moments of time that could be 
used to define the partition between these phases were 
introduced and discussed. Firstly the saturation of the 
average $\tau_{sat}$ was considered followed by the cut 
off moment $\tau_{cutoff}$. Then we noted that if one 
restricts the interactions to be between nearest neighbours, 
the entanglement initially scales as the area of the smaller
region and in the second phase as that of the volume. This 
led us to introduce $\tau_{vol}$, the moment the entanglement 
is typically volume scaling. Of these perhaps the cut off 
moment $\tau_{cutoff}$ is the most attractive choice, since 
it gives an unambiguous single time that corresponds to the 
moment that the entanglement probability distribution is 
equal to, for practical purposes, that of the unbiased 
distribution of pure states.

The results support the relevance of generic entanglement 
as we show it can be generated efficiently from two-qubit 
gates. Therefore protocols relying on typical/generic 
entanglement, like
\cite{abeyesinghe, harrow, buhrman, hayden} gain relevance\cite{oliveira}.
 
The above results may be extended in various directions 
that will be described briefly here.

{\em Multi-particle entanglement measures --} It would 
be interesting to extend the above results to further investigate properties 
of typical multi-partite entanglement. For entanglement measures 
based on average purities \cite{Plenio V 05} this is
possible with the results established here. Other measures 
such as the entanglement of formation may also be treated 
but require extensions of the approach presented here that 
go beyond the scope of this paper.

{\em Markov Process Quantum Monte Carlo Methods --} Our 
results possess another interpretation that is of interest 
for the numerical study of quantum-many-body systems. One 
numerical approach to classical spin systems is to evolve 
spin configurations randomly according to the Metropolis 
rule, i.e. always accepting moves that decrease energy and 
accepting moves that increase energy with a probability
proportional to $exp\{-\Delta E/kT\}$. This reproduces 
the correct thermal average. One may of course consider 
a similar approach in the quantum setting applying random 
two-qubit gates to progress the state. Our present results 
fall into this category but apply for infinite temperatures 
as we draw our unitaries from the invariant Haar measure. 
However, we may adapt our analysis to the finite temperature 
regime employing stabilizer states at the expense of having 
to analyze a more complicated Markov process. Basic ideas 
of our approach carry over and open the possibility for 
rigorous statements concerning the convergence rate of such 
a Markov Process Quantum Monte Carlo approach.

{\em Continuous Variables --} These question are made 
additionally complicated in the continuous variable setting 
through the lack of a Haar measure in the setting of 
non-compact groups. However in \cite{serafini} an approach 
to proceed is introduced. There one notes that it is 
reasonable to assume that the maximum energy of the 
global pure state is finite. This tames the non-compactness 
and one can define a way to pick states at random. 
Reference \cite{serafini,serafini2} gives an explicit method to achieve that for Gaussian states, which probably will take 
on the role analogous to stabilizer states in the general 
setting. This opens up the possibility of studying the 
questions dealt with here in the continuous setting.

{\em Two phases --} The relation between the three
moments of time that can be used to separate the two 
phases we observe should be further investigated. We 
believe analytical tools for studying the cut-off effect 
are sufficiently developed to undertake an analysis 
with the aim of proving relations between $\tau_{vol}$, 
$\tau_{cutoff}$, and $\tau_{sat}$ respectively.

{\em Relation to other work --} It would be interesting 
to investigate how our results relate to existing work 
on Spin-gases\cite{Briegel} which are similarly semi-quantal 
systems. We also hope to relate this line of enquiry
to work touching on what the typical entanglement of 
the universe is\cite{popescu,gemmer}. The main difference 
to the approach here would be to consider closed systems.

{\em Experimental studies --}
Optical lattices appear to provide a suitable experimental 
setting to test the results obtained here. There is also 
hope of using Bose-Einstein condensates or linear 
optics to study the continuous variable case. 

{\em Acknowledgements --}
We gratefully acknowledge initial discussion with 
Jonathan Oppenheim as well as discussions with
Koenraad Audenaert, Fernando Brandao, Benoit Darquie, 
Jens Eisert, David Gross, Konrad Kieling, Terry Rudolph, 
Alessio Serafini, Graeme Smith, John Smolin, Andreas Winter, and Aram
 Harrow.
 
We acknowledge support by the EPSRC QIP-IRC, The Leverhulme 
Trust, EU Integrated Project QAP, the Royal Society, the NSA, 
the ARDA through ARO contract number W911NF-04-C-0098, 
and the Institute for Mathematical Sciences at Imperial 
College London.\newline \newline

\section{Appendix A: Uniform (Haar) measure on the unitary group}\label{sec:Haar}
We here briefly introduce the often used uniform measure on pure states, 
sometimes called the unitarily invariant measure. This is a particular instance of a Haar measure, which can be viewed as a generalization of 
the 'flat distribution'. A flat probability distribution is the one reflecting no bias with respect to any object. 

Such a distribution on a set of objects is invariant under permutations of the said objects. Therefore if there 
is a group of transformations associated with the set of elements, the distribution on those elements should be invariant under application of
elements of the group. The simplest non-trivial example is probably a coin. Here there are two group elements, the identity and the flip.
Only the unbiased probability distribution P(Head)=1/2 and P(Tail)=1/2 is invariant under those transformations.

\begin{figure}[th]
\centerline{
\includegraphics[width=5cm]{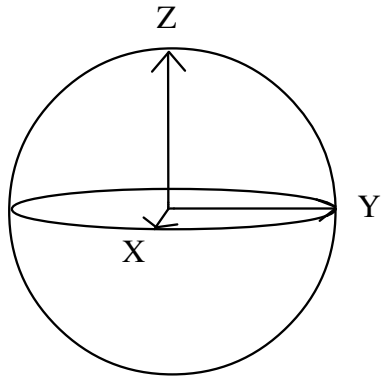}
}
\vspace{.0cm}
\caption{Bloch sphere: the unitarily invariant distribution corresponds to a uniform distribution over this sphere.}
\label{fig:bloch} 
\end{figure}

Bearing this in mind, consider picking pure general states at random. The unbiased distribution on pure states is here required to be
 invariant under unitary transforms, i.e. $P(\ket{\Psi})=P(U\ket{\Psi})$, where it is implicit that we are in a continuous setting. This requirement uniquely defines the distribution. 
For a single qubit this can be nicely visualised as a uniformly dense distribution on the Bloch sphere, see figure \ref{fig:bloch}.

The normal method to pick pure states from this distribution is to fix an arbitrary pure state and apply a unitary picked at random from the associated measure on
unitary matrices. See for example \cite{diaconisrm} for the explicit procedure.

\section{Appendix B: Randomizing properties of Haar measure}\label{sec:assumption}

 The following lemma shows that Requirement \ref{assump:rand} is satisfied by our example with Haar measure.

\begin{lemma}\label{lem:allweneed}Suppose $F=F(A,B)$ is a bi-linear function of one-qubit operators $A$, $B$, that $\sigma_a,\sigma_b$ are two Pauli matrices, and that $I$ is randomly drawn from Haar measure on $U(2)$. Then
$$\Ex{F(T\sigma_aT^\dag,T \sigma_b T^\dag)} = \left\{\begin{array}{ll}F(I,I),& a=b=0;\\ \frac{1}{3}\sum_{w\in\{x,y,z\}}F(\sigma_w,\sigma_w), & a=b\neq 0\\ 0, &\mbox{otherwise.}\end{array}\right.$$\end{lemma}

To prove it, we need an intermediate result.

\begin{lemma}\label{lem:sigmas}Assume that $\sigma_a$, $a\neq 0$, is a Pauli operator and $T$ is a random unitary drawn from Haar measure on $U(2)$. Then
$$T \sigma_a T^\dag = r_x\sigma_x + r_y\sigma_y + r_z\sigma_z,$$
where $(r_x,r_y,r_z)$ is a random vector whose distribution is invariant under permutations. Moreover, $\Ex{r_{u}r_{u'}}=\delta_{u,u'}/3$ for all $u,u'\in\{x,y,z\}$.\end{lemma}

\begin{proof}$T \sigma_a T^\dag$ is a $2\times 2$ Hermitian matrix. Therefore, there exist real numbers (in fact random variables) $r_u = \Tr(\sigma_uT \sigma_a T^\dag)/2$ ($u\in\{0,x,y,z\}$) such that
$$T \sigma_w T^\dag = r_0 I + r_x\sigma_x + r_y\sigma_y + r_z\sigma_z.$$
$\Tr{(T \sigma_a T^\dag)} = \Tr(\sigma_a)=0$ implies $r_0=0$. 
Before we continue, let us state a simple fact we will use repeatedly.

\begin{claim}[{Conjugation trick}] For any unitary $Q\in U(2)$, $T \sigma_a T^\dag$ and $QT\sigma_aT^\dag Q^\dag = QT\sigma_a(QT)^\dag$ have the same probability distribution. Therefore, the distribution of $(r_x,r_y,r_z)$ is the same as that of $(r'_x,r'_y,r'_z)$, where 
\begin{equation}\label{eq:conjtrick}r'_u =\frac{1}{2}\, \Tr(\sigma_uQT\sigma_wT^\dag Q^\dag)=\frac{1}{2}\,\Tr((Q\sigma_uQ^\dag)T\sigma_wT^\dag).\end{equation}\end{claim}

In fact, this follows directly from the invariance property of Haar measure. Since $T$ and $QT$ have the same probability distribution, so do $T \sigma_a T^\dag$ and $QT\sigma_a(QT)^\dag$ (which are the same deterministic function of $Q$ and $QT$, respectively) and the same holds for $(r_x,r_y,r_z)$ is the same as that of $(r'_x,r'_y,r'_z)$. 

We now apply the trick as follows.

\begin{enumerate}
\item Take $Q = H$ (the Hadamard matrix). Then $Q\sigma_xQ^\dag = \sigma_z$, $Q\sigma_z Q^\dag=\sigma_x$ and $Q\sigma_y Q^\dag = \sigma_y$. This implies that $2r'_x = Tr((Q\sigma_xQ^\dag)T\sigma_aT^\dag) = Tr(\sigma_yT\sigma_wT^\dag)=2r_y$, $2r'_y = Tr((Q\sigma_yQ^\dag)T\sigma_aT^\dag) = Tr(\sigma_zT\sigma_aT^\dag)=2r_x$ and $r'_z=r_z$. Hence $(r'_x,r'_y,r'_z)=(r_z,r_y,r_x)$ and $(r_x,r_y,r_z)$ have the same distribution.
\item Now take $Q=\ketbra{0} - i\ketbra{1}$. $Q\sigma_xQ^\dag = \sigma_y$, $Q\sigma_y Q^\dag=\sigma_x$ and $Q\sigma_zQ^\dag = \sigma_z$. It follows from the above reasoning that $(r_y,r_x,r_z)$ and $(r_x,r_y,r_z)$ have the same distribution. 
\item Take $Q=\sigma_z$ this time. Then $Q\sigma_xQ^\dag = - \sigma_x$, $Q\sigma_yQ^\dag = -\sigma_y$, $Q\sigma_zQ^\dag=\sigma_z$, so $(-r_x,-r_y,r_z)$ and $(r_x,r_y,r_z)$ have the same distribution. Similarly 
Similarly, we can take $Q=\sigma_y$ or $Q=\sigma_x$ to show that $(-r_x,r_y,-r_z)$ and $(r_x,-r_y,-r_z)$ also have the same distribution.
\end{enumerate}
The first two items show that the distribution of $(r_x,r_y,r_z)$ is invariant by transposition of the $x$ and $z$ coordinates and of the $x$ and $y$ coordinates. It follows that the distribution is also invariant under transposition of the $y$ and $z$ coordinates (which is a composition of a $xz$ transposition followed by a $xy$ transposition and another $xz$ transposition). Since any permutation is a composition of transpositions, we have shown that the the distribution of $(r_x,r_y,r_z)$ is invariant under permutations of the coordinates. Moreover, it also follows that
$$1 = \frac{\Tr(\sigma_w^2)}{2} = \frac{1}{2}\Ex{\Tr[(T\sigma_wT^\dag)^2]}=\Ex{r_x^2+r_y^2+r_z^2} = 3\Ex{r_{u}^2}\mbox{ for all }u\in\{x,y,z\}.$$
Thus it only remains to show that $\Ex{r_{u}r_{u'}}=0$ if $u\neq u'$. To this end, we use item $3.$ If for instance $u=x$, $u'=z$ we recall that $(-r_x,-r_y,r_z)$ and $(r_x,r_y,r_z)$ have the same distribution, hence $r_xr_z$ and $-r_xr_z$ also have the same distribution, which implies our claim. The other cases follow similarly.\end{proof}
We can now prove \lemref{allweneed}.

\begin{proof}First assume that $a\neq b$. Without loss of generality, assume that $b\neq 0$. We apply an idea based on the Conjugation Trick from the previous proof. There exists $c\in\{x,y,z\}$ such that $\sigma_a$ and $\sigma_c$ commute and $\sigma_b$ anti-commutes with $\sigma_c$. As $U$ and $U\sigma_c$ have the same distribution, $(U\sigma_c)\sigma_a(U\sigma_c)^\dag=U\sigma_aU^\dag$ and $(U\sigma_c)\sigma_b(U\sigma_c)^\dag=-U\sigma_bU^\dag$, we have
\begin{eqnarray*}\Ex{F(U\sigma_aU^\dag,U\sigma_bU^\dag)} &=& \Ex{F((U\sigma_c)\sigma_a(U\sigma_c)^\dag,(U\sigma_c)\sigma_b(U\sigma_c)^\dag)}\\
&=& -\Ex{F(U\sigma_aU^\dag,U\sigma_bU^\dag)},\end{eqnarray*}
thus the expected value is $0$.
Now if $a=b=0$, the result is trivial. If $a=b\neq 0$, we can write $U\sigma_aU^\dag = r_x\sigma_x + r_y\sigma_y + r_z\sigma_z$ as in \lemref{sigmas}, and by bi-linearity
\begin{equation}\Ex{F(U\sigma_aU^\dag,U\sigma_aU^\dag)} = \sum_{w',w''}\Ex{r_{w'}r_{w''}}F(\sigma_{w'},\sigma_{w''}).\end{equation}
Applying \lemref{sigmas} finishes the proof.\end{proof}

\section{Appendix C: Stabilizer states}
\noindent

Stabilizer states are a discrete subset of general quantum states,
which can be described by a number of parameters scaling
polynomially with the number of qubits in the state
\cite{gottesman1,gottesman2,NC}.

A {\em stabilizer operator} on $N$ qubits is a tensor product of
operators taken from the set of Pauli operators
\begin{equation}
\sigma_1:=\left(\begin{array}{cc}0&1\\1&0\end{array}\right),\,\,
\sigma_2:=\left(\begin{array}{cc}0&-i\\i&0\end{array}\right),\,\,
\sigma_3:=\left(\begin{array}{cc}1&0\\0&-1\end{array}\right),
\end{equation}
and the identity $I$. An example for $N=3$ would be the operator
$g=\sigma_1 \otimes I \otimes sigma_3$. A set $G=\{g_1,\ldots,g_K\}$ of $K$
mutually commuting stabiliser operators that are independent,
i.e.\ $\prod_{i=1}^{K} g_i^{s_i}=I$ exactly if all $s_i$ are
even, is called a {\em generator set}. For $K=N$ a generator
set $G$ uniquely determines a single state $|\psi\rangle$ that
satisfies $g_k|\psi\rangle=|\psi\rangle$ for all $k=1,\ldots,N$.
Such a generating set generates the stabilizer group.
Each unique such group in turn defines a unique {\em stabilizer state}.

For example the GHZ state $\ket{000}+\ket{111}$ is defined
by the generator set $\langle \sigma_1\otimes\sigma_1\otimes\sigma_1, \sigma_0\otimes\sigma_3\otimes\sigma_3, \sigma_3\otimes\sigma_3\otimes\sigma_0 \rangle$.

A key observation that is useful for the 
considerations here is the fact that the bipartite entanglement of a
stabilizer state, i.e. the entanglement across any bipartite split, takes only integer
values \cite{Audenaert P 05,fattal}.

Finally we note the fact that in order for the stabilizer state to be non-trivial it
is necessary and sufficient that the elements of the stabilizer group (a) commute,
and (b) are not equal to {\em -I} \cite{NC}.

\end{document}